%% file: paper.tex
\pdfoutput=1
\documentclass[journal]{vgtc}                % final (journal style)
\usepackage[utf8]{inputenc}

\ifpdf%                                % if we use pdflatex
  \pdfoutput=1\relax                   % create PDFs from pdfLaTeX
  \usepackage{graphicx}                % allow us to embed graphics files
  \DeclareGraphicsExtensions{.pdf,.png,.jpg,.jpeg} % for pdflatex we expect .pdf, .png, or .jpg files
\else%                                 % else we use pure latex
  \usepackage{graphicx}                % allow us to embed graphics files
  \DeclareGraphicsExtensions{.eps}     % for pure latex we expect eps files
\fi%

\graphicspath{{figures/}{pictures/}{images/}{./}} % where to search for the images

\usepackage{microtype}                 % use micro-typography (slightly more compact, better to read)
\PassOptionsToPackage{warn}{textcomp}  % to address font issues with \textrightarrow
\usepackage{textcomp}                  % use better special symbols
\usepackage{mathptmx}                  % use matching math font
\usepackage{times}                     % we use Times as the main font
\usepackage{cite}                      % needed to automatically sort the references
\usepackage{tabu}                      % only used for the table example
\usepackage{booktabs}                  % only used for the table example
\onlineid{0}

\vgtccategory{Research}
\vgtcpapertype{algorithm/technique}

\usepackage{cellspace}

\usepackage{ifpdf}

\ifpdf%                                % if we use pdflatex
  \pdfoutput=1\relax                   % create PDFs from pdfLaTeX
  \usepackage{graphicx}                % allow us to embed graphics files
  \DeclareGraphicsExtensions{.pdf,.png,.jpg,.jpeg} % for pdflatex we expect .pdf, .png, or .jpg files
\else%                                 % else we use pure latex
  \usepackage{graphicx}                % allow us to embed graphics files
  \DeclareGraphicsExtensions{.eps}     % for pure latex we expect eps files
\fi%

\usepackage{tabularx}

\graphicspath{{figures/}{pictures/}{images/}{./}}

\usepackage{microtype}         
\PassOptionsToPackage{warn}{textcomp} 
\usepackage{textcomp}                 
\usepackage{mathptmx}                
\usepackage{times}                    
       
\usepackage{cite}                   
\usepackage{tabu}                   
\usepackage{booktabs}               
\usepackage[normalem]{ulem} 
\usepackage{xspace}
\usepackage{amsfonts}
\usepackage{amsmath}
\usepackage{multirow}  % For table

\usepackage[utf8]{inputenc}
\usepackage[T1]{fontenc}
\usepackage{balance}
\usepackage[hyphens]{url}
\usepackage{tabularx}

\usepackage{wrapfig,epsf}

\usepackage{xcolor}

\usepackage[hidelinks]{hyperref}

\newcommand{\ie}{\emph{i.e.}\xspace}
\newcommand{\eg}{\emph{e.g.}\xspace}

\newcommand{\myparagraph}[1]{\smallskip\noindent\textbf{#1}\xspace}

\newcommand{\figref}[1]{\hyperref[#1]{\textbf{Figure~\ref*{#1}}}}
\newcommand{\figsref}[1]{\hyperref[#1]{Figures~\ref*{#1}}}
\newcommand{\tabref}[1]{\hyperref[#1]{Tab.~\ref*{#1}}}
\newcommand{\secref}[1]{\hyperref[#1]{Sec.~\ref*{#1}}}
\newcommand{\algref}[1]{\hyperref[#1]{Alg.~\ref{#1}}}

\title{LegalVis: Exploring and Inferring Precedent Citations in Legal Documents}

\author{Lucas~E.~Resck*, Jean~R.~Ponciano*, Luis~Gustavo~Nonato~\textit{Member,~IEEE}, Jorge~Poco~\textit{Member,~IEEE}}

\authorfooter{
\item[*] The first two authors contributed equally to this work.
\item Lucas~E.~Resck and Jean~R.~Ponciano are with Funda\c{c}\~ao Getulio Vargas, Brazil. E-mail: lucas.domingues@fgv.edu.br, jean.ponciano@fgv.br
\item Jorge Poco is with Funda\c{c}\~ao Getulio Vargas, Brazil and Universidad Católica San Pablo, Perú. E-mail: {jorge.poco@fgv.br}.
\item Luis Gustavo Nonato is with ICMC-USP, S\~ao Carlos, Brazil. E-mail: {gnonato@icmc.usp.br}.
}

\abstract{
To reduce the number of pending cases and conflicting rulings in the Brazilian Judiciary, the National Congress amended the Constitution, allowing the Brazilian Supreme Court (STF) to create \textit{binding precedents} (BPs), \ie, a set of understandings that both Executive and lower Judiciary branches must follow. The STF's justices frequently cite the 58 existing BPs in their decisions, and it is of primary relevance that judicial experts could identify and analyze such citations.
To assist in this problem, we propose LegalVis, a web-based visual analytics system designed to support the analysis of legal documents that cite or could potentially cite a BP.
We model the problem of identifying potential citations (\ie, non-explicit) as a classification problem. However, a simple score is not enough to explain the results; that is why we use an interpretability machine learning method to explain the reason behind each identified citation.
For a compelling visual exploration of documents and BPs, LegalVis comprises three interactive visual components: the first presents an overview of the data showing temporal patterns, the second allows filtering and grouping relevant documents by topic, and the last one shows a document's text aiming to interpret the model's output by pointing out which paragraphs are likely to mention the BP, even if not explicitly specified.
We evaluated our identification model and obtained an accuracy of 96\%; we also made a quantitative and qualitative analysis of the results.
The usefulness and effectiveness of LegalVis were evaluated through two usage scenarios and feedback from six domain experts.
}

\input{figures.tex}

\vgtcinsertpkg

\begin{document}


\maketitle

\input{1-intro.tex}

\input{2-related.tex}
\input{3-overview.tex}

\input{4-data.tex}

\input{5-model.tex}

\input{6-visual.tex}
\input{7-cases.tex}
\input{8-experts.tex}
\input{9-discussion.tex}

\input{10-conclusion.tex}

% \balance
% \clearpage

 \acknowledgments{
This work was supported by CNPq-Brazil (grants \#303552/2017-4 and \#312483/2018-0), Rio de Janeiro State Funding Agency (FAPERJ)-Brazil (grant \#E-26/201.424/2021), S\~ao Paulo Research Foundation (FAPESP)-Brazil (grant \#2013/07375-0), and the School of Applied Mathematics at Fundação Getulio Vargas (FGV). 
Any opinions, findings, conclusions, or recommendations expressed in this material are those of the authors and do not necessarily reflect the views of the CNPq, FAPESP, FAPERJ, or FGV.
}

\bibliographystyle{abbrv-doi}

\bibliography{paper}
%\balance
 
%\vspace*{-10px}

\vbox{%
\begin{wrapfigure}{l}{70pt}
{
%\vspace*{-20px}
	\includegraphics[width=1in,height=1.25in,clip,keepaspectratio]{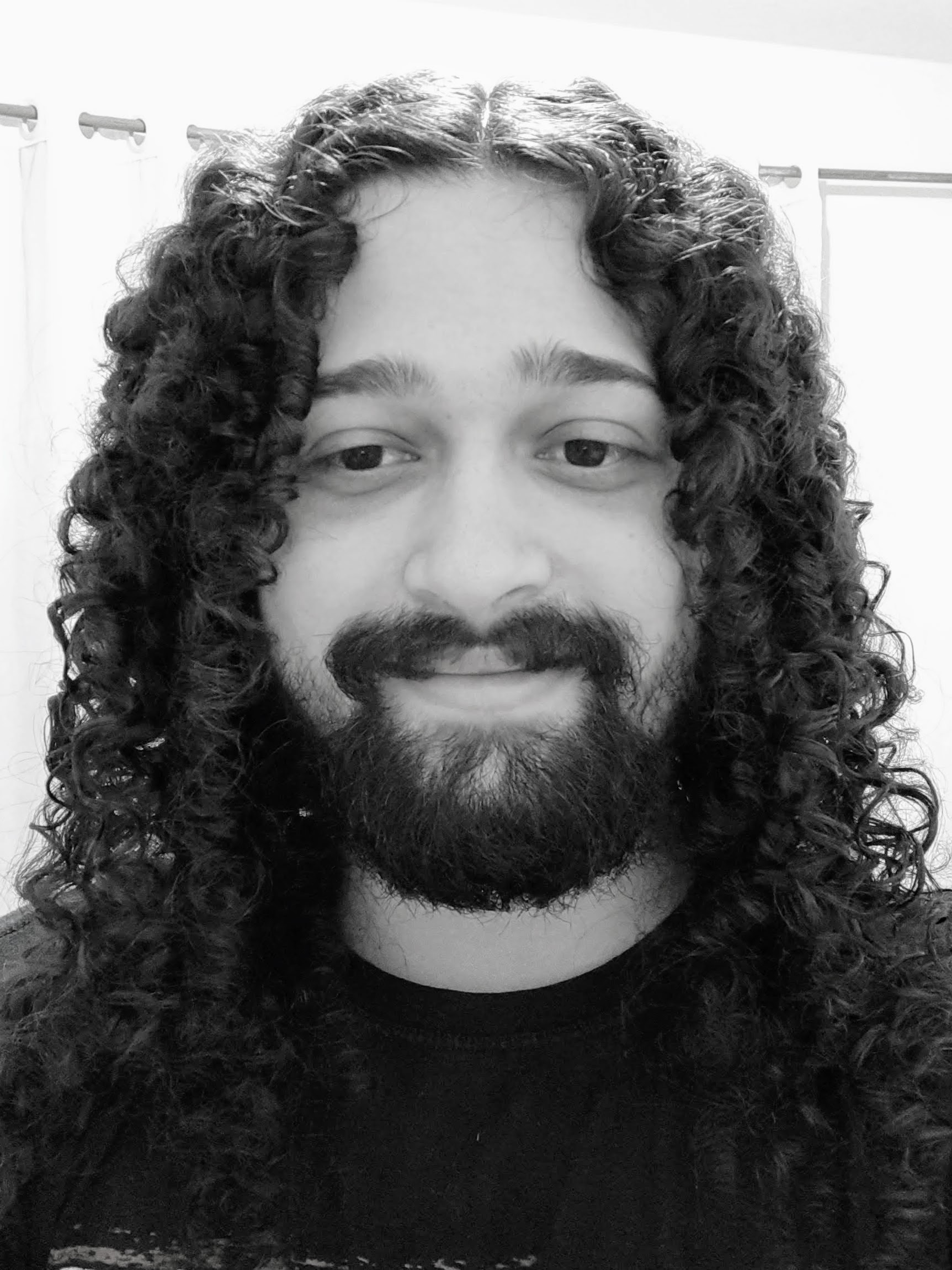}
	%\vspace*{15pt}
}%
\end{wrapfigure}
\noindent\small 
\\
\\
\textbf{Lucas~E.~Resck}
  received his BSc (2021) in Applied Mathematics from the School of Applied Mathematics of Fundação Getulio Vargas (FGV-EMAP), Rio de Janeiro, Brazil, where he is currently a MSc student in Mathematical Modeling.
  His research interests include machine learning, natural language processing, and machine learning explainability.}

\vbox{%
\begin{wrapfigure}{l}{70pt}
{
\vspace*{25pt}
	\includegraphics[width=1in,height=1.25in,clip,keepaspectratio]{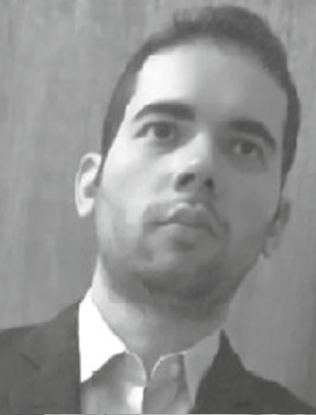}
%	\vspace*{15pt}
}%
\end{wrapfigure}
\noindent\small 
\\
\\
\\
\\
\textbf{Jean~R.~Ponciano}
is currently a postdoctoral fellow at the School of Applied Mathematics of Fundação Getulio Vargas, Brazil. He received
the BSc (2012), MSc (2016), and Ph.D. (2020) degrees in computer science from the Federal University of Uberlandia, Brazil. His research interests include information visualization, visual analytics, network science, and data streams. He has served as a program committee member or external reviewer in relevant conferences, including EuroVis and ASONAM.}

\vbox{%
\begin{wrapfigure}{l}{70pt}
{
\vspace*{20pt}
	\includegraphics[width=1in,height=1.25in]{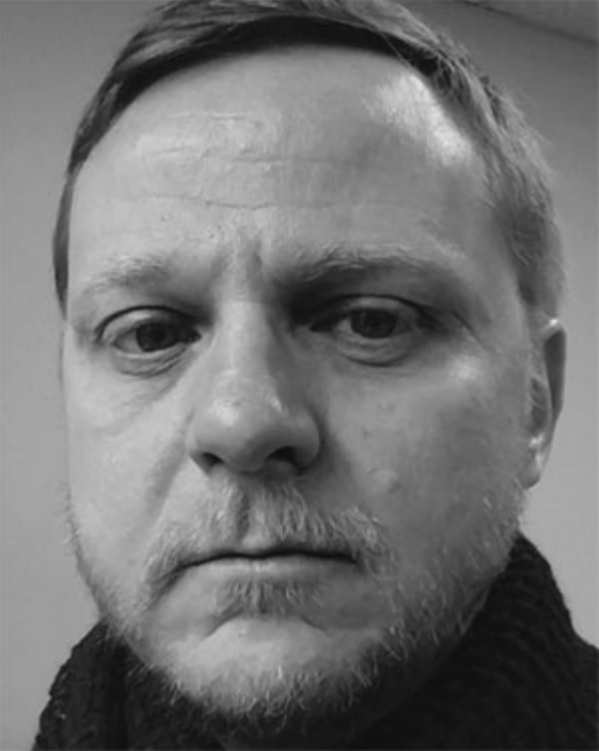}
%	\vspace*{15pt}
}%
\end{wrapfigure}
\noindent\small 
\\
\\
\\
\textbf{Luis~Gustavo~Nonato}
received a Ph.D. degree in applied mathematics from the Pontificia Universidade Catolica do Rio de Janeiro, Rio de Janeiro-Brazil, in 1998. His research interests include visualization, visual analytics, machine learning, and data science. He is a full professor with the Institute of Mathematical and Computer Sciences, University of S\~ao Paulo, S\~ao Carlos, Brazil. He was a visiting professor in the Center for Data Science, New York University, New York, from 2017 to 2018. From 2008 to 2010, he was a visiting scholar in the Scientific Computing and Imaging Institute, University of Utah, Salt Lake City. Besides serving in several program committees, including IEEE SciVis, IEEE InfoVis, and EuroVis, he was associate editor of the Computer Graphics Forum and IEEE TVCG. He was also the editor-in-chief of the SBMAC SpringerBriefs in Applied Mathematics and Computational Sciences.}

\vbox{%
\begin{wrapfigure}{l}{70pt}
{
\vspace*{13pt}
	\includegraphics[width=1in,height=1.25in,clip,keepaspectratio]{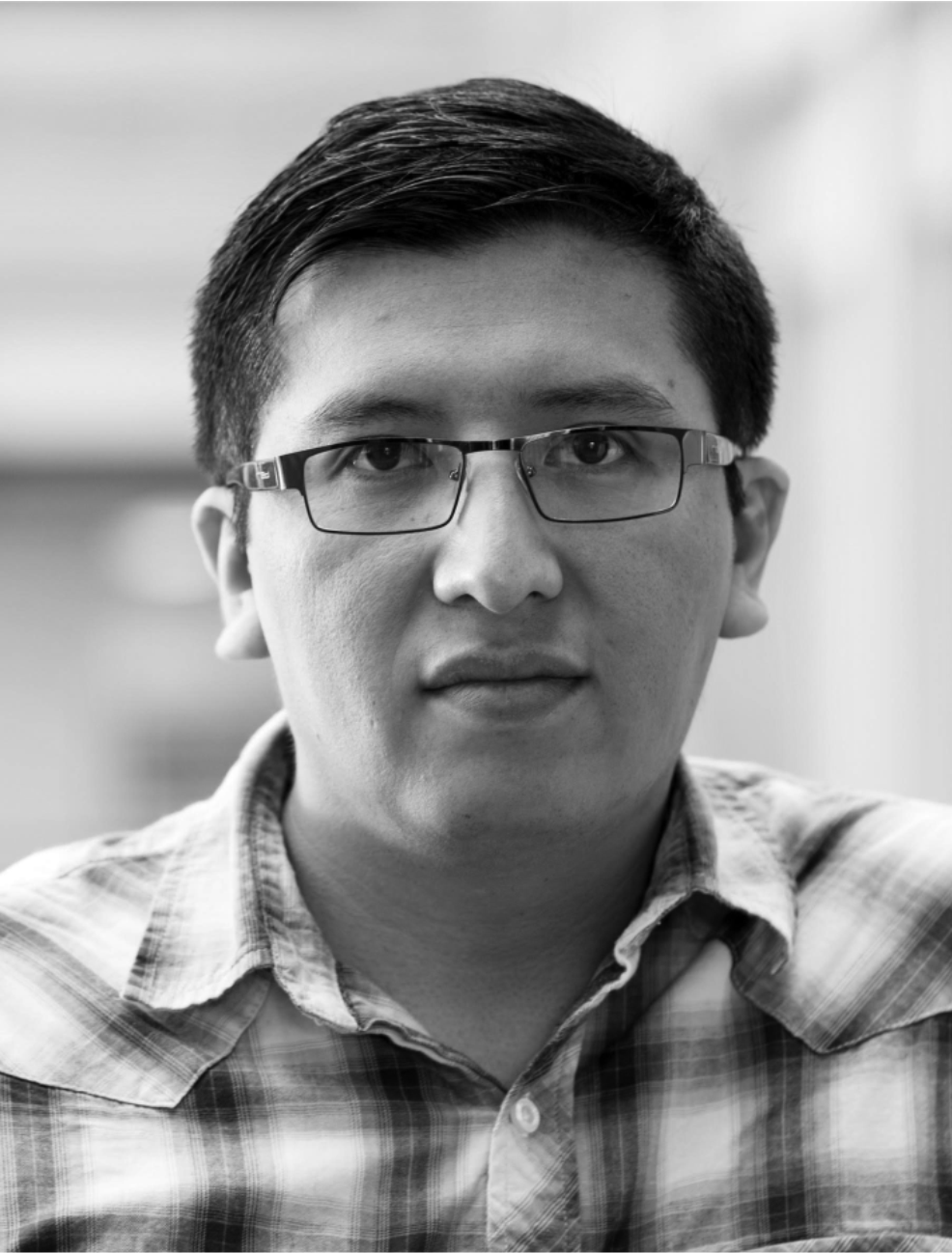}
%	\vspace*{15pt}
}%
\end{wrapfigure}
\noindent\small 
\\
\\
\\
\textbf{Jorge~Poco}
is an Associate Professor at the School of Applied Mathematics at Fundação Getulio Vargas Rio de Janeiro-Brazil. He received his Ph.D. in Computer Science in 2015 from New York University, his M.Sc. in Computer Science in 2010 from the University of São Paulo (Brazil), and his B.E. in System Engineering in 2008 from National University of San Agustín (Peru).
His research interests are data visualization, visual analytics, machine learning, and data science. He has served in several program committees, including IEEE SciVis, IEEE InfoVis, VAST, and EuroVis.}

\end{document}

\maketitle

\begin{strip}\centering
\vspace{-2cm}
\includegraphics[width=\linewidth]{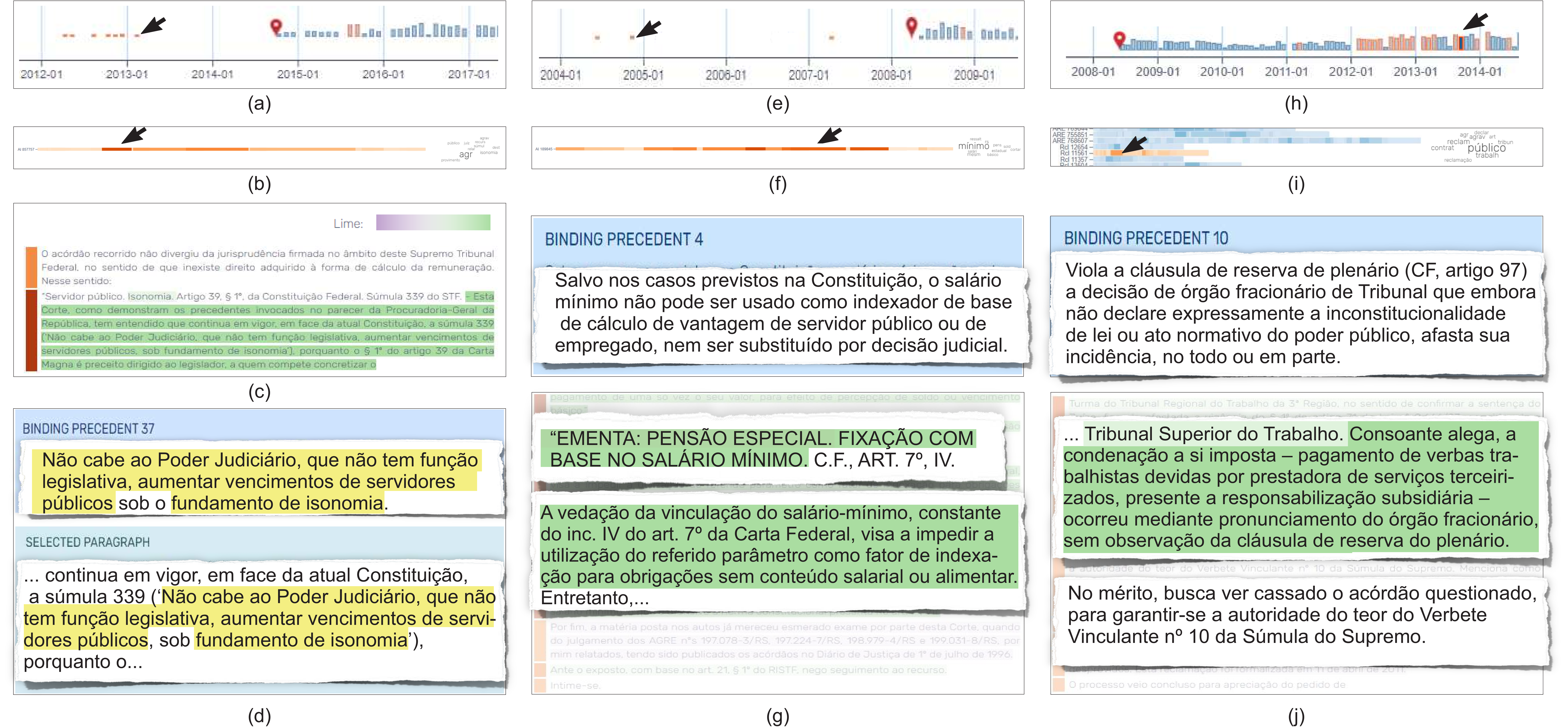}

\captionof{figure}{Three potential citations to different binding precedents -- original texts (in Portuguese) shown. (left) a document with the entire BP 37's text but published before creating this binding precedent. (middle) a document published before the creation of BP 4 that potentially cites its underlying reasoning. (right) a document potentially citing BP 10 while adopting a different terminology to refer to this binding precedent (unlabeled document).
\label{fig:supcaseStudy1}}
\end{strip}

\section{Original Images in Portuguese}

Through two usage scenarios (Sect.~7) validated by domain experts, our paper demonstrates the usefulness of LegalVis to assist users in exploring legal documents and binding precedents. In the first usage scenario, we focused on highlighting the  system's capabilities in identifying potential citations under different situations. Here, we show in~\figref{fig:supcaseStudy1} the original texts (in Portuguese) of those parts of the binding precedents and documents explored throughout this scenario. Our second usage scenario, by its turn, showed an exploratory analysis that aimed to find patterns and behaviors related to particular justices and types of documents. To help in the analysis, we presented parts of a decision reported by justice Edson Fachin (Fig.~8~(d)). Here,~\figref{fig:supcaseStudy2} shows the original texts (in Portuguese) that compose those parts.

\figSupMatCaseTwo

\newpage

\section{Word Count Histogram}

\figref{fig:histogram} shows a word count histogram of the documents considered in our study (i.e., 29,743 documents that explicitly cite at least one of the ten most cited BPs plus 30,000 documents without
explicit citations). In general, we have less than four thousand words.
\figHistogram

%% file: figures.tex
\newcommand*{\img}[1]{%
    \raisebox{-.3\baselineskip}{%
        \includegraphics[
        height=\baselineskip,
        width=\baselineskip,
        keepaspectratio,
        ]{#1}%
    }%
}

\newcommand{\figWorkflow}{
\begin{figure*}[ht!]
  \centering
  \includegraphics[width=\linewidth]{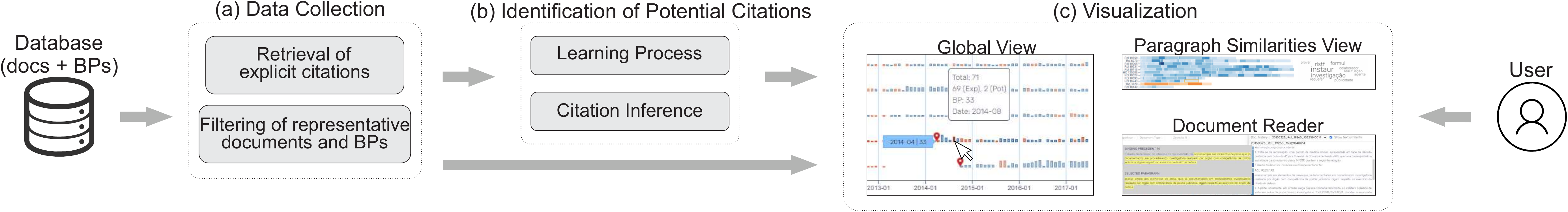}
  \caption{LegalVis workflow. (a) we first retrieve explicit citation information and filter representative decisions/BPs from our database. (b) these data are used in an elaborate and two-fold strategy to infer potential citations. Finally, (c) the user can interactively explore documents that cite or could potentially cite binding precedents through the three visual components that compose our framework.}
\label{fig:workflow}
\end{figure*}
}

\newcommand{\figCitations}{
\begin{figure}[t!]
  \includegraphics[width=0.93\linewidth]{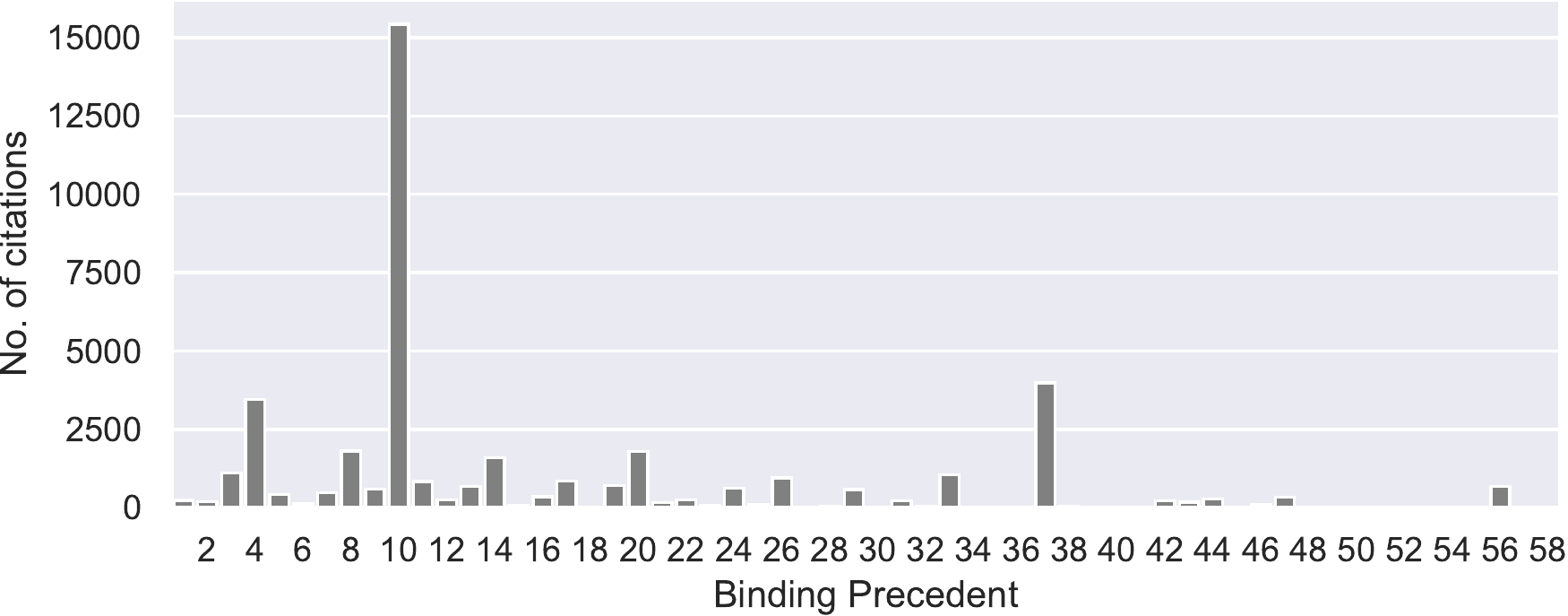}
  \caption{Number of documents with a BP citation. Note that many BPs have just a few citations.}
  \label{fig:citations_bp}
\end{figure}
}

\newcommand{\figPipeline}{
\begin{figure}[t!]
  \centering
  \includegraphics[width=\linewidth]{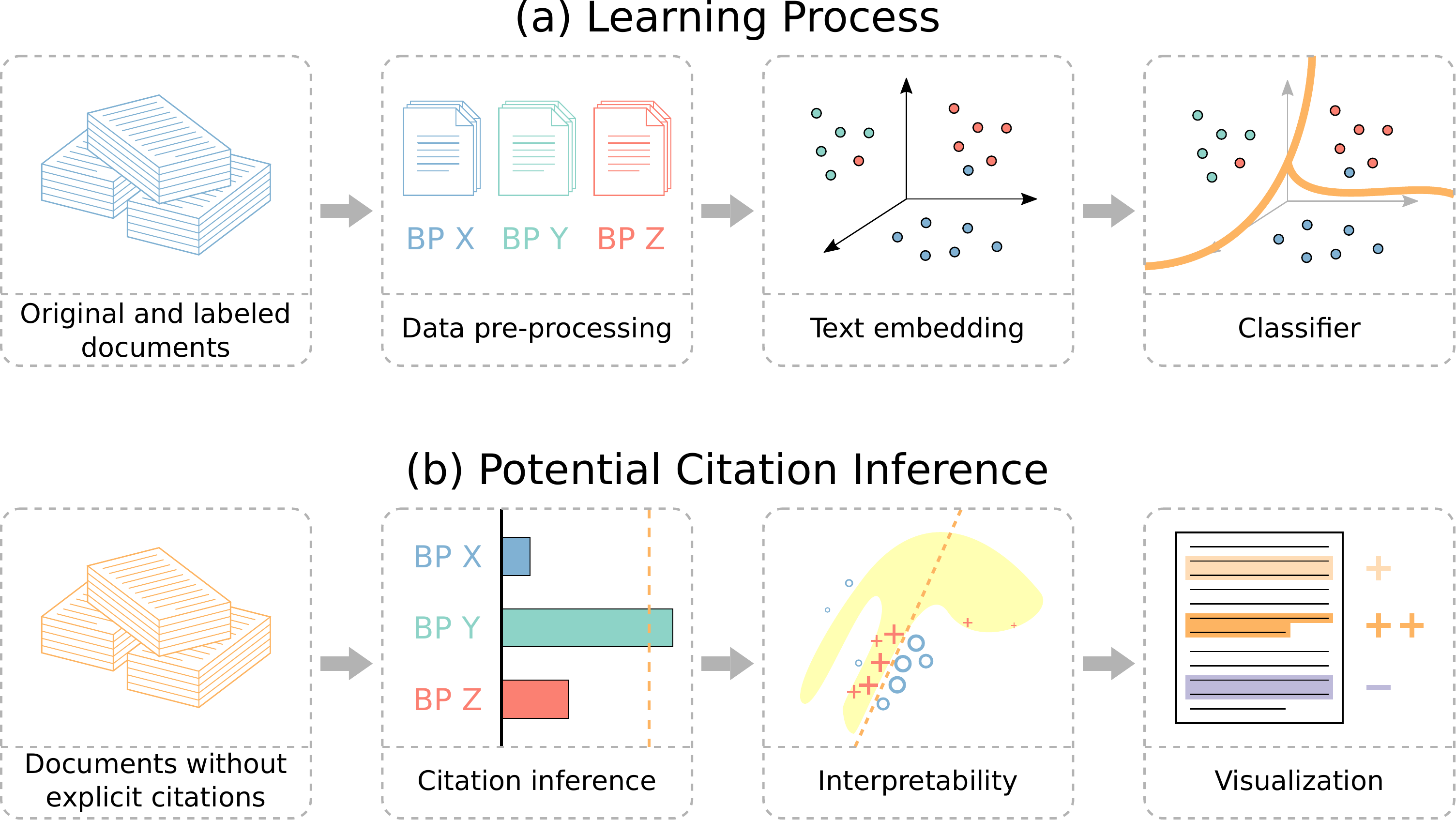}
  \caption{Pipeline of identification of potential citations. (a) Labeled documents are pre-processed and used to fit a classifier that learns what characterizes a citation. (b) The fitted classifier is used to infer potential citations for documents without explicit citations, and an interpretability technique explains the inference, presenting the most relevant sentences.}
\label{fig:potential_pipeline}
\end{figure}  
}

\newcommand{\figConfusion}{
\begin{figure}[t!]
  \centering
  \includegraphics[width=0.8\columnwidth]{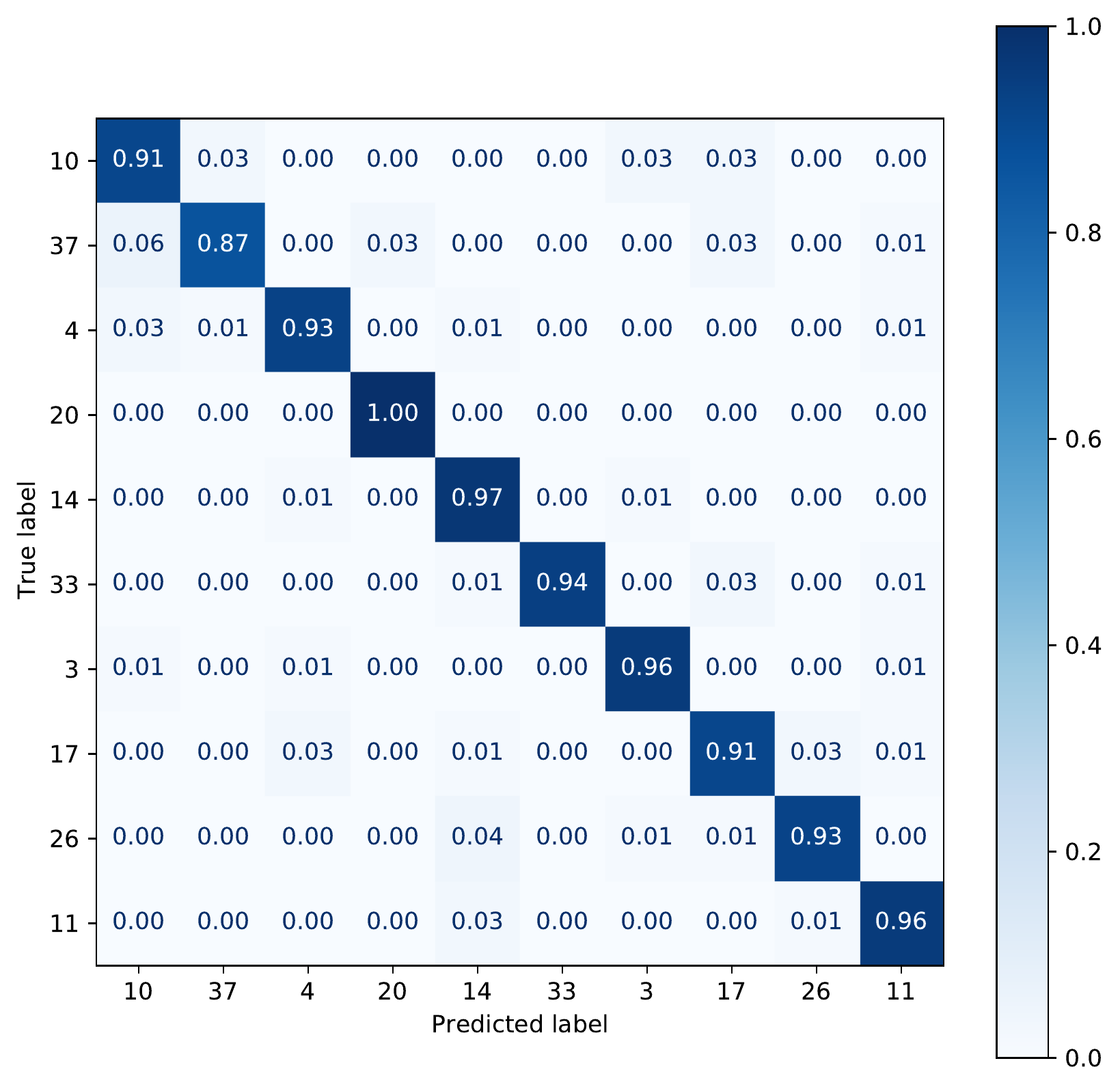}
  \caption{Confusion matrix of Truncated TF-IDF with linear SVM on validation portion.} 
  \label{fig:confusion_matrix}
\end{figure}
}

\newcommand{\figVisualComp}{
\begin{figure*}[t!]
  \centering
  \includegraphics[width=\linewidth]{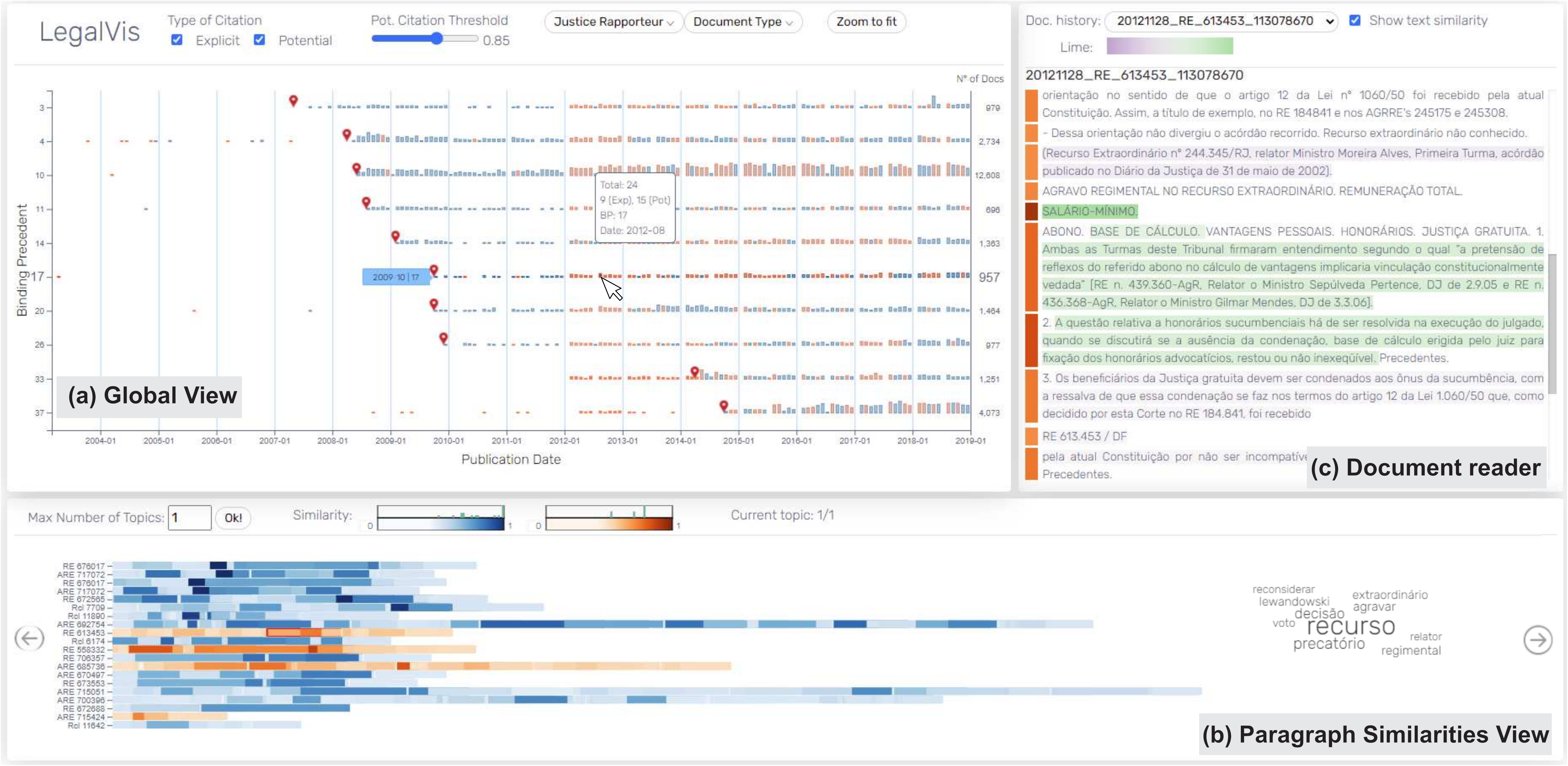}
  \caption{LegalVis is composed of three interactive views designed to support the exploration of documents and binding precedents: \textit{Global View}, \textit{Paragraph Similarities View}, and \textit{Document Reader}.}
  \label{fig:visualComponents}
  \end{figure*}
}

\newcommand{\figShowTextSimilarity}{
\begin{figure}[t!]
  \centering
  \includegraphics[width=\linewidth]{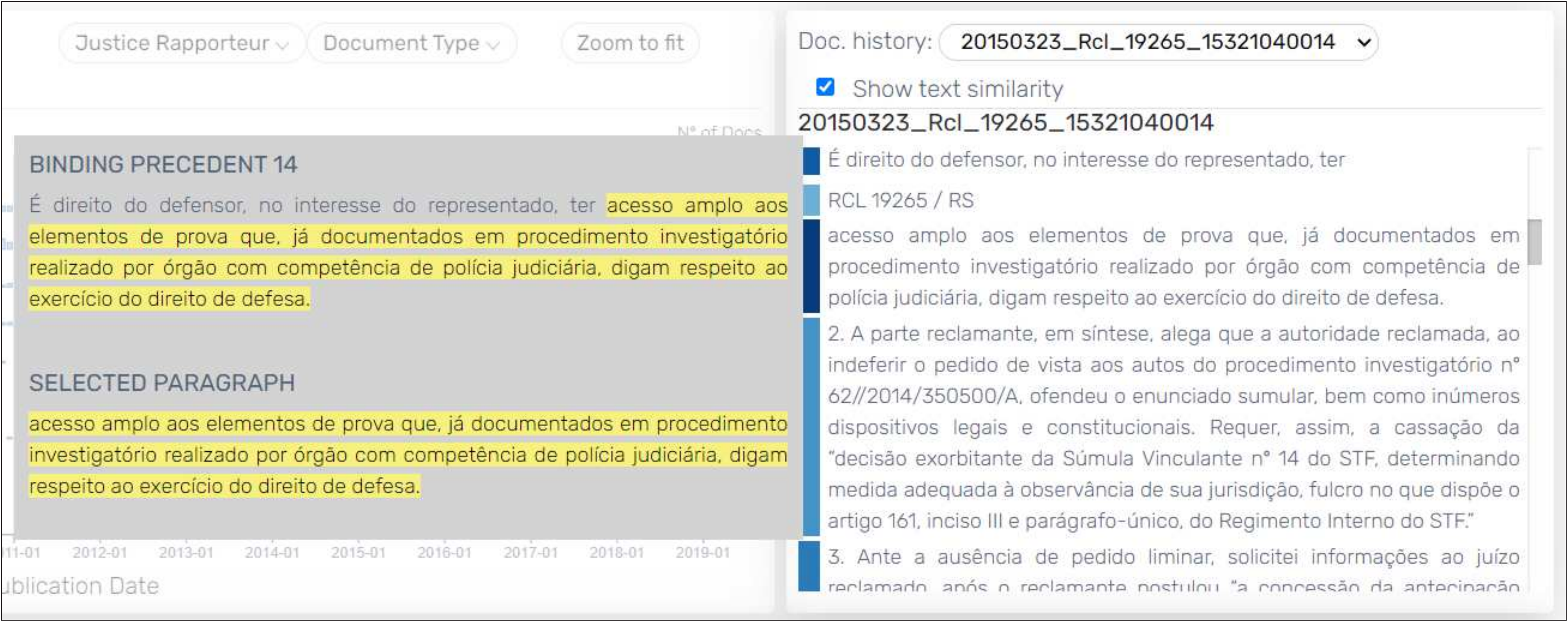}
  \caption{Comparison between a selected paragraph and the binding precedent. By hovering over the paragraphs of a selected document in Document Reader, users can compare their contents with the binding precedent being cited. Optionally, common parts are highlighted to help in the analysis (yellow).}
  \label{fig:show_text_similarity}
\end{figure}
}

\newcommand{\figCaseOne}{
\begin{figure*}[t!]
  \centering
  \includegraphics[width=\linewidth]{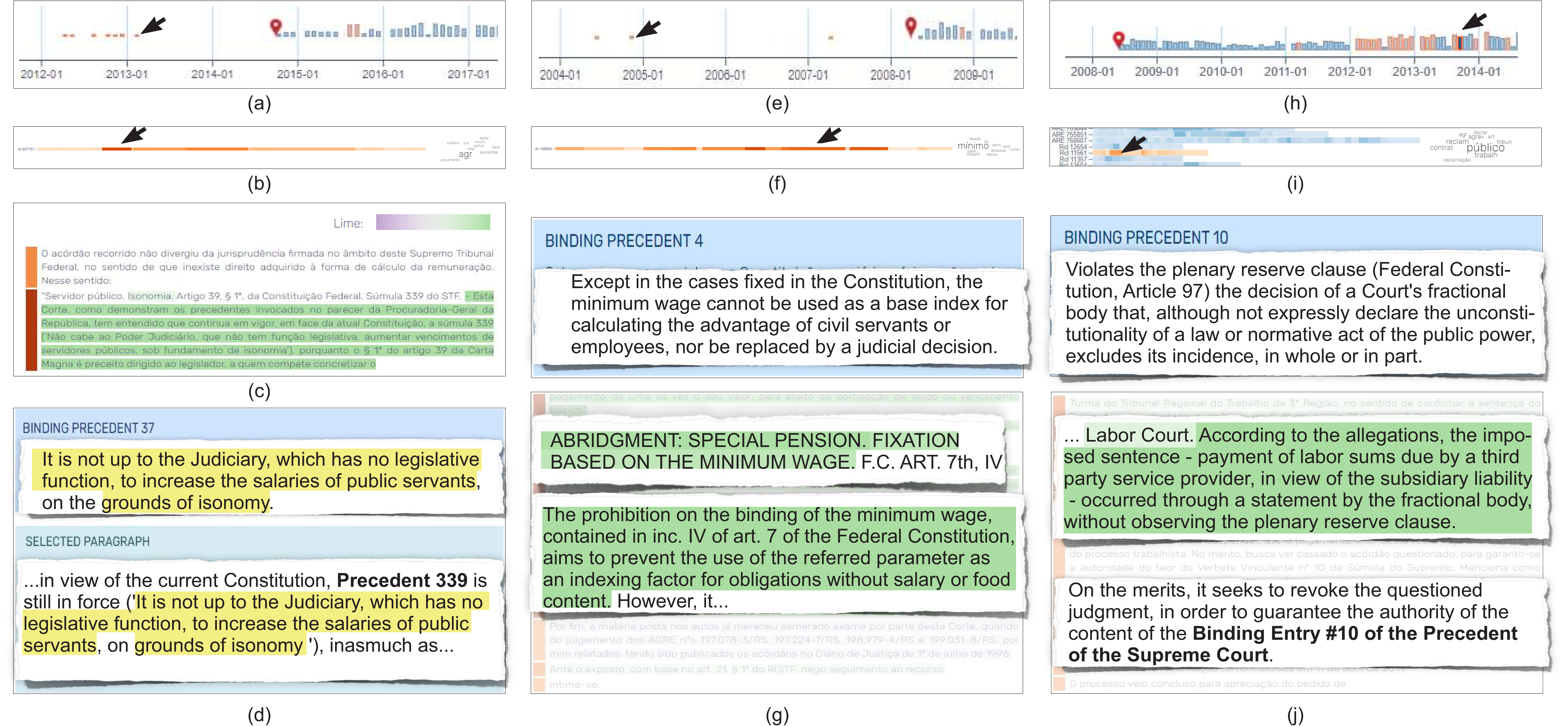}
  \caption{Three potential citations to different binding precedents. (left) a document with the entire BP 37's text but published before creating this binding precedent. (middle) a document published before the creation of BP 4 that potentially cites its underlying reasoning. (right) a document potentially citing BP 10 while adopting a different terminology to refer to this binding precedent (unlabeled document).}
  \label{fig:caseStudy1}
\end{figure*}
}

\newcommand{\figCaseTwo}{
\begin{figure}[t!]
  \centering
  \includegraphics[width=\linewidth]{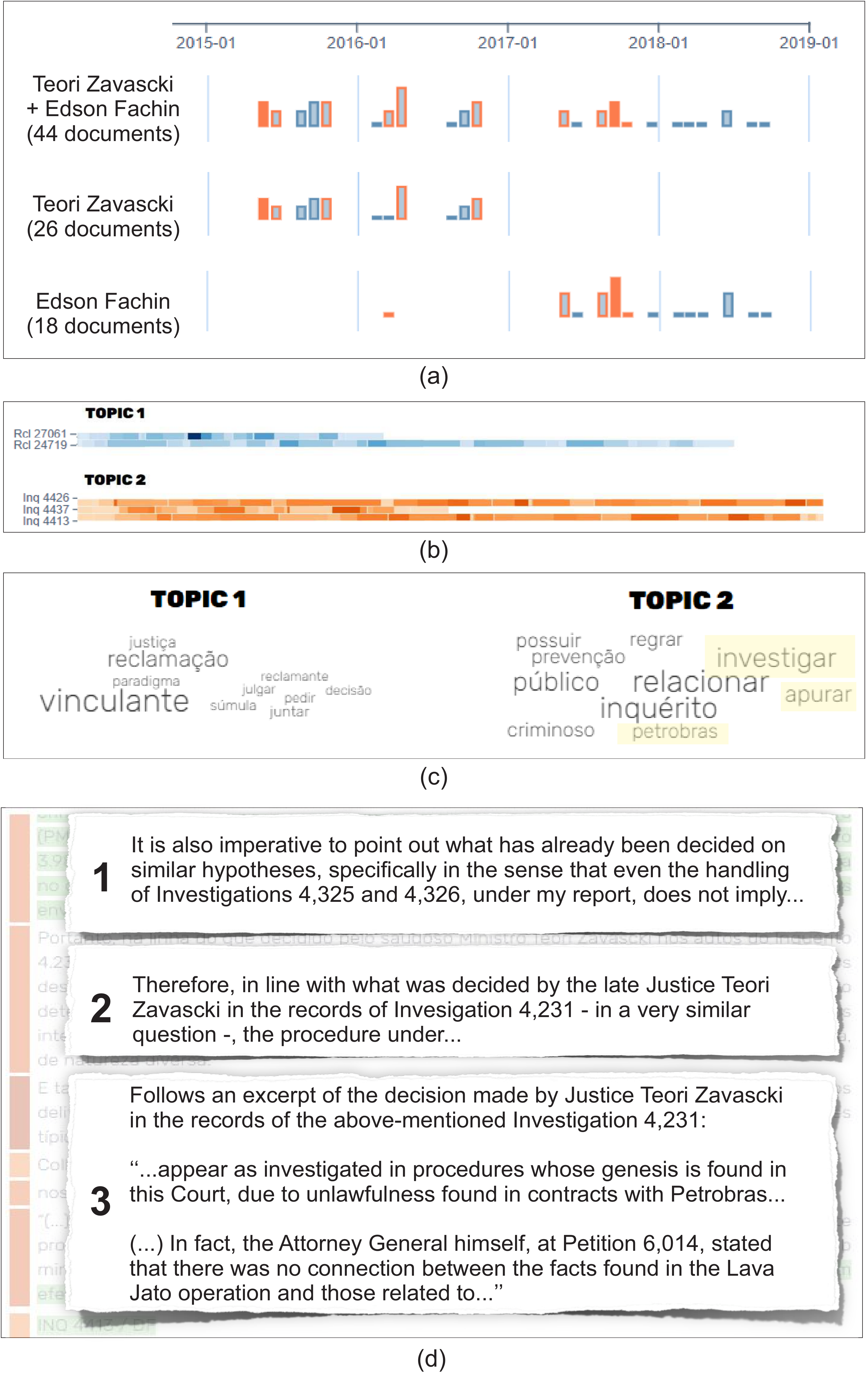}
  \caption{Exploratory analysis showing a decision reported by Edson Fachin that is related to other reported by Teori Zavascki on the same subject: Lava Jato operation.}
  \label{fig:caseStudy2}
\end{figure}
}

\newcommand{\figLikert}{
\begin{figure}[t!]
  \centering
  \includegraphics[width=1\linewidth]{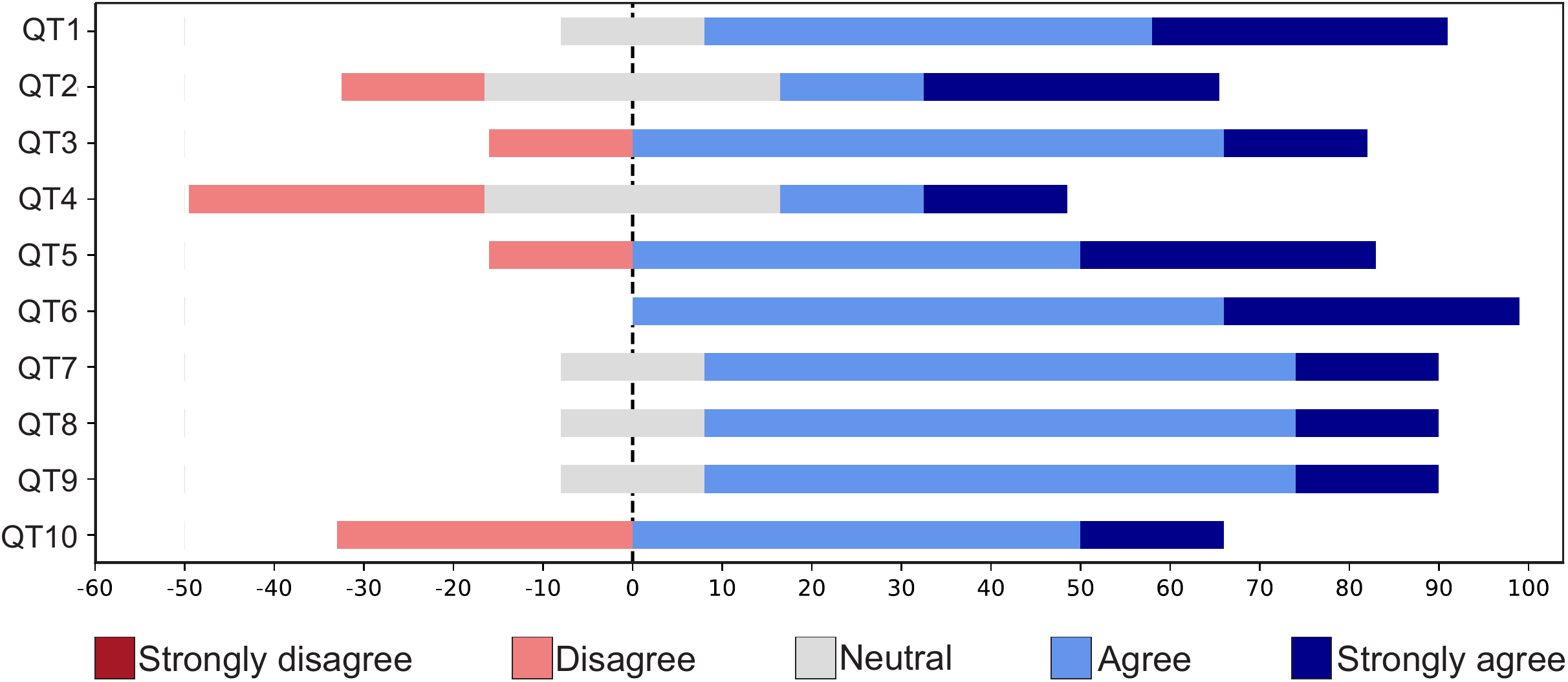}
  \caption{Answers for questions QT1-QT10 according to six domain experts not involved in the tool's development. The bar length indicates the percentage of respondents who have chosen that specific Likert level.}
  \label{fig:likert}
\end{figure}
}

%% file: 1-intro.tex
\section{Introduction}\label{sec:introduction}

The Brazilian Supreme Court (\textit{Supremo Tribunal Federal} (STF) in Portuguese) is the highest court of law in Brazil. It is primarily responsible for guarding the rights in the Brazilian Constitution. 
There are eleven justices, nominated by the President and confirmed by the Senate. 
Among its multiple attributions, STF can provide a final judgment on appeals from other courts and a judicial review on some norms created in the country.

In 2004, Brazil had more than 100 million pending cases in the Judiciary\cite{cnjpendingcases}. 
To reduce those numbers and avoid conflicting decisions, the National Congress amended the Constitution that same year, allowing STF to create binding precedents (Súmulas Vinculantes in Portuguese). 
These precedents consolidate understandings on judicial issues that both Executive and lower Judiciary branches must follow. 
From 2004 to 2020, the Court ruled 58 of those precedents, covering some of the most critical discussions from crime to tax law\cite{stfsumulas}. For example, binding precedent 37 addresses the issue of salaries of public servants and has the following statement:
\textit{``It is not up to the Judiciary, which has no legislative function, to increase the salaries of public servants, on the grounds of isonomy.''}
At first, the intention was to end the controversy. Still, the STF ends up later resolving thousands of cases that present a divergence on the application of a binding precedent (BP).
Consequently, STF's justices regularly cite these binding precedents in their decisions. 

We consider a decision as a justice's ruling on a particular legal case, written and published on a legal document afterward. The terms ``decision'' and ``legal document'' are used interchangeably in this paper.
Given the large number of decisions in STF\,---\,more than 1 million from 2011 to 2020 \cite{stfnumberrulings}\,---\,, lawyers and other judicial experts face difficulties finding citations to BPs and analyzing them.
Such difficulties arise from the lack of a computational tool to help the experts to find, for instance, decisions that cite a particular BP of interest. 
Moreover, it is usual to have decisions with hundreds of pages, making it hard for lawyers and other experts to find parts of interest where justices cite, quote, or implicitly mention a BP.

We worked alongside a domain expert to address these challenges to develop a web-based visual analytic tool named LegalVis. 
We designed this system to explore and analyze the texts of STF's legal documents, particularly their explicit or implicit relationships with binding precedents.
In contrast to an explicit citation to a BP, which can be easily found by searching ``Binding Precedent \#X'' (in free translation) in the document's content, finding potential citations to a BP is not a straightforward task. 
To tackle this issue, LegalVis relies on a machine learning model to identify potential citations, building upon an interpretability mechanism to provide reliable explanations for the model's decisions.
This system also provides mechanisms for quickly identifying documents and BPs of interest through multiple linked views and interactive tools. 
These features permit easy analysis of the content of the documents and, at the same time, allow us to interpret the relationships that exist between similar documents and their relation with the BPs.
We demonstrate the potential of LegalVis through two usage scenarios that show insightful and relevant findings.
Our usage scenarios were validated by six domain experts not involved in the system's development, who also outlined the usefulness and potential of LegalVis.

Both components in LegalVis, the potential citation identification model and the visual analytics tool, could be applied to any court within the judicial system. We have chosen to deal with STF's decisions due to their relevance in the national context. 
Our main contributions are:

\begin{itemize}
  \item A pipeline for the identification of potential (\ie, non-explicit) citations that relies on (i) a machine learning model and (ii) an interpretability mechanism that provides reliable explanations for the model's decisions. Properly handling legal data, especially non-English data, is a sensitive task by its nature. Our pipeline and obtained results benefit this application domain and may inspire other applications.
  \item LegalVis, a visual analytics system that assists lawyers and other judicial experts in exploring legal documents and binding precedents.
  \item Two usage scenarios, validated by six domain experts not involved in the system's development, showing relevant judicial findings concerning STF's decisions.
\end{itemize}

%% file: 2-related.tex
\section{Related Work}

In this paper, we are interested in identifying and analyzing decisions somehow related to binding precedents, either by explicitly citing or implicitly mentioning them. In this context, a crucial step is to compute and visualize the similarity between the binding precedent and parts of a legal document. In the following, we discuss related research under two perspectives: visualization of similarities between parts of documents and works related to legal documents analysis.

\subsection{Visualizing Similarities between Parts of Documents}

Several text visualization techniques have been proposed for many tasks, domains, and data types. We point the reader to the surveys by Alharbi and Laramee~\cite{alharbi2018sos} and by Kucher and Kerren~\cite{kucher2015text} for a broader understanding of this topic.

One particular category of techniques focuses on visualizing document-level similarity, which is generally achieved by representing documents as points in a 2D or 3D visualization plane and positioning them accordingly~\cite{cao2016overview, alsakran2011streamit, van2008visualizing}.
Rather than just knowing whether two documents are similar to each other, one may be interested in exploring similarity at a fragment-level, \ie, by visually comparing parts (or fragments) of documents (\eg, paragraphs or sentences) to a reference text and highlighting their similarities.

A high similarity between the two parts is expected when there is text repetition, which is one of the text reuse possibilities~\cite{surveyTextAlignment}. Still, even parts that do not share many words may be similar to each other depending on the underlying semantics~\cite{use}. Either way, a visual analysis of these similar parts' contents is helpful in various contexts, as in tasks related to technical writing~\cite{soto2015similarity} and plagiarism identification~\cite{potthast2013overview,riehmann2015visual}. Examples include \textit{PicaPica}~\cite{riehmann2015visual}, a visual analytic tool that highlights differences and commonalities between two documents to help users in detecting plagiarism, and the \textit{Text Re-use Browser}~\cite{janicke2014visualizations}, which highlights pairs of similar sentences of the Bible via a dot plot matrix.

Different visualization techniques support the comparison between documents at a fragment-level, from 
grid-based heatmaps~\cite{janicke2014visualizations, abdul2017constructive} to side-by-side views~\cite{riehmann2015visual}. They usually fall into one of two basic categories depending on the level of details on the raw text they provide~\cite{janicke2017visual}: while \emph{close reading} techniques allow the analysis of the text itself (words, phrases, ideas, and so on\,---\,see, \eg, \textit{Text Re-use Reader}~\cite{janicke2014visualizations}), \emph{distant reading} techniques\footnote{\emph{Close reading} and \emph{distant reading} are terms firstly created to categorize visualization techniques applied to digital humanities~\cite{cheema2016annotatevis}.} generate abstract representations that summarize the texts' information (see, \eg, \textit{Compare Cloud}~\cite{diakopoulos2015compare} and ViTA~\cite{ abdul2017constructive}). As stated by Jänicke et al.~\cite{janicke2017visual}, hybrid strategies that guide users from distant to close reading (top-down approach) successfully apply the well-established visual information-seeking mantra \emph{``overview first, zoom and filter, then details-on-demand"}~\cite{visualmantra}. Visual analytic tools usually rely on linked views to accomplish hybrid visualizations. 
One such example is the system proposed by Kiesel et al. to compare argument structures in essays~\cite{kiesel2020visual}.

Our approach computes syntactic and semantic similarities (for explicit and potential citations, respectively) between parts of a document and the binding precedent being cited. To show such similarities intuitively and effectively, we rely on a hybrid approach that allows both distant and close readings in a top-down manner. Contrary to the mentioned systems, LegalVis is designed to handle Brazilian legal documents and particularities. We discuss studies related to the analysis of legal documents in the following section.

Finally, different visual encoding strategies may be employed to optimize the analysis of similar parts' contents when in close reading. One such choice is the use of colors (in the background or font color) to indicate common words or shared ideas~\cite{riehmann2015visual}. Other possibilities include changes in the font size, connections, and glyphs~\cite{janicke2017visual}. LegalVis employs the first option.

\subsection{Legal Documents Analysis}

In common law countries, such as the United States and Australia, lawyers and other judicial experts often search for precedents, looking at how old decisions could help solve open cases~\cite{zhang2007semantics}. In this context, automated text processing becomes a valuable tool that allows these professionals to search and analyze legal documents~\cite{barros2018case,legal_document_review}, especially when the task of interest involves finding relevant precedents or similar decisions~\cite{correia2019exploratory, mandal2017measuring}.

Given the many benefits of visual analytics, some authors have created visualization tools to help users explore such a high amount of legal data. An example is \textit{Knowlex}~\cite{lettieri2017legal}, a visual analytic system that links and exhibits Italian legal documents from different sources related to the user's law of interest. By employing citation networks, \textit{EUCaseNet}~\cite{lettieri2016computational} allows one to visually explore citations involving judgments from the European Court of Justice through network structural properties that will enable, \eg, recognizing relevant precedents and judges' behaviors. Visualizations applied to other countries' legal data also exist, for example, in Netherlands~\cite{wyner2017answering} and Portugal~\cite{carvalho2018transforming}.

Despite advances in the development of visual frameworks to assist in legal documents analysis, users are still challenged in exploring and visualizing legal data in Brazil. The Brazilian courts' websites and other official platforms usually exhibit copious pages of results, each containing several and long decisions\cite{barros2018case}. A few efforts have been made to enhance the visualization of these data. One of them is the \textit{Supremo 2.0} system~\cite{chada2015visualizing}, a tool designed to visualize STF's quantitative and multidimensional data. By working with a dataset similar to ours, Gomez-Nieto et al.~\cite{gomez2015understanding} proposed a visual analytic tool that allows exploring case-related aggregated information from STF through different representations (\eg, stacked graphs, treemaps, and heatmaps).

Our approach addresses a challenge not pursued by any other tool, even though some of them have characteristics in common with ours, such as linked views or the use of STF's data. LegalVis's primary goal is to provide a compelling exploration of legal documents that explicitly or potentially cite binding precedents. This involves an effective visualization and suitable machine learning and interpretability methods to identify and explain potential (non-explicit) citations. Not even those studies that also rely on STF's data~\cite{gomez2015understanding,chada2015visualizing} consider this type of citation information in their work.

\myparagraph{Other tools and search engines.} Various platforms developed search engines for legal documents and jurisprudence.
In Brazil, \textit{Buscador Dizer o Direito}~\cite{dizerDireito} presents commented documents, although the number of commentaries may be considered small, and \textit{OAB Juris~}\cite{oabjuris} uses artificial intelligence algorithms to present more relevant results. Other Brazilian platforms include \textit{JusBrasil}~\cite{jusbrasil} and \textit{LexML}~\cite{lexml}.
In the Netherlands, \textit{Bluetick}~\cite{bluetick} offers suggestions and similarity concepts during the search.
In general, however, these platforms do not offer efficient visualization or natural language processing (NLP) algorithms that could improve results and time during a user's search.
Some other tools are more elaborate, offering solutions based on analytics, visualization, machine learning, digitalization, and process automation. In Brazil, \textit{Finch Platform}~\cite{finchplatform} and \textit{Legal One}~\cite{legalone} are some of the most known tools, and, in United States, there are \textit{Lexis}~\cite{lexis}, \textit{Westlaw}~\cite{westlaw}, and \textit{Casetext}~\cite{casetext}. The last one, in particular, allows the use of Transformer-based models~\cite{transformer} to search for case laws that are similar to given sentences. As stated, none of these tools and search engines meets LegalVis's objectives.

%% file: 3-overview.tex
\section{System Overview}

After several weekly meetings with a domain expert, we identified the challenges behind exploring legal documents and binding precedents.
We characterize a set of questions/requirements to be addressed by the analytical tool. 
We present the system's requirements and the concrete visualization tasks that guided our visual design throughout this section. 
We also provide an overview of the system's workflow.

\figWorkflow

\subsection{System Requirements}

After elucidating the domain experts' needs, we came up with several questions grouped into three categories. 

The first category comprises questions related to the identification of decisions that cite or could potentially cite a binding precedent:

\myparagraph{Q1.1.} Which decisions in the STF (overall, per Justice rapporteur, and type of decision) explicitly cite a specific binding precedent?

\myparagraph{Q1.2.} Which decisions in the STF (overall, per Justice rapporteur, and type of decision) could potentially cite a specific binding precedent? In other words, which decisions end up citing a particular binding precedent in a non-explicit manner?

\myparagraph{Q1.3.} Are there decisions related to a to-be-created binding precedent?

\vspace{0.2cm}

The second category of questions are related to the filtering and analysis of relevant decisions:

\myparagraph{Q2.1} Which decisions cite a specific binding precedent on a given date?

\myparagraph{Q2.2.} Given a date, are there similar decisions that cite the same binding precedent?

\myparagraph{Q2.3.} Which are the most relevant decisions among those that cite a specific binding precedent on a given date regarding the similarity between binding precedent and decision?

Finally, the third category of questions addresses issues related to the similarity between parts of decisions and BPs and the relevance of those parts:

\myparagraph{Q3.1.} How similar is each part of a decision to the corresponding binding precedent?

\myparagraph{Q3.2.} Which parts of a decision cite or are likely to cite a specific binding precedent?

\myparagraph{Q3.3.} Given many decisions, can one quickly access multiple decisions of interest and their relevant parts when analyzing different dates and binding precedents?

\subsection{Design Tasks}
Based on the requirements described above, we raised concrete visualization tasks to guide the LegalVis system's development, following the mantra \emph{``overview first, zoom and filter, then details-on-demand"}~\cite{visualmantra}.

\myparagraph{T1 - Identification of explicit and potential citations:} The system should exhibit decisions that explicitly cite a specific binding precedent (Q1.1) and identify/exhibit those that could potentially mention it (non-explicit citation) (Q1.2, Q1.3).

\myparagraph{T2 - Overview of decisions and binding precedents:} The system should provide an overview of all the decisions and binding precedents, highlighting the chronological order they appear (Q1.3).

\myparagraph{T3 - Filtering and selecting decisions:} The system should allow filtering decisions by Justice rapporteur, by decision type, and by type of citation (explicit or potential) (Q1.1, Q1.2, Q1.3). It should also allow selecting decisions based on the date and binding precedent (Q2.1).

\myparagraph{T4 - Grouping and ordering decisions:} The system should identify and group similar decisions that cite the same binding precedent on a given date (Q2.2) and order them in the layout according to the similarity between binding precedent and parts of a decision (Q2.3).

\myparagraph{T5 - Highlight in decisions' relevant parts:} The system should quickly identify the most similar paragraphs/sentences to the binding precedent (Q3.1). It should also highlight those parts that (either explicitly or potentially) cite that precedent (Q3.2).

\myparagraph{T6 - Document browsing history:} The system must provide means to access decisions already analyzed quickly, regardless of the date and binding precedent (Q3.3).

\subsection{Workflow}

After mapping the requirements into design tasks, we define the system's workflow (see \figref{fig:workflow}). 
In the first stage, \textbf{data collection (a)}, we consult the digitized decisions stored in a database. 
They already contain the explicit citations that were found using regular expressions. 
However, it is necessary to query and label the decisions and binding precedents that we will use in the following stages (details in \secref{sec:data_set}).

In the second stage, \textbf{identification of potential citations (b)}, we use the collected data to train a machine learning model that characterizes a citation to a particular binding precedent.
After that, the procedure infers potential citations and explains the reason behind the decision by employing an interpretability method (details in \secref{sec:potentialCitation}).

Finally, in the third stage, \textbf{visualization (c)}, users explore and visualize documents that explicitly cite or that can potentially be citing binding precedents. 
LegalVis comprises three linked view components:
Global View, which presents an overview of the data under a temporal perspective;
Paragraph Similarities View, which allows for filtering and grouping relevant documents; and
Document Reader, which shows a document's content and points out which parts of the document are likely to mention a binding precedent (details in \secref{sec:visual}).

%% file: 4-data.tex
\section{Dataset}
\label{sec:data_set}

\figCitations

When an STF justice decides a case, this decision is consolidated in a document. 
Each document contains the decision's text, publication date, the justice that conducted the process (justice rapporteur), and the document type (\eg, Complaint or Extraordinary Appeal). 
In this work, we are interested in STF decisions that are somehow related to a binding precedent, explicitly or \emph{potentially} citing it.

In total, we have 58 binding precedents, idealized as a mechanism to create a consolidated understanding among the STF's justices. 
We collected our dataset from a partnership with the \emph{Supremo em Números} (``STF in Numbers'', in free translation) project~\cite{falcao2013relatorio}\,---\,a project that seeks to assess legal and computer knowledge to produce unprecedented data on the Supreme Court. 
This dataset contains more than 2,500,000 documents since 1988 and metadata such as the number of the BP being cited (if that is the case) and document type.
Considering only decisions that explicitly cite at least one of the 58 BPs, there are 38,364 documents, totaling 41,031 citations (some documents may mention more than one BP). 
The number of citations per BP does not follow a uniform distribution, as we can see in \figref{fig:citations_bp}.

As requested by the domain experts, the name of the justice rapporteur is essential for the analysis. %
However, this information was not in the initial dataset, so we extracted it using regular expressions in the document's content. 
As part of the data cleaning process, we had some difficulties: (i) documents with different titles but same content (we refer to them as \emph{duplicated} documents), and (ii) documents in which it is not trivial to extract out the justice name.
However, these problems did not impact our work's development, and we give more details in~\secref{sec:limitations}.

In this work, we decided to deal with documents that explicitly cite at least one of the ten most-cited BPs (\ie, 3, 4, 10, 11, 14, 17, 20, 26, 33, and 37).
We made this decision to have a representative sample of documents and citations. In fact, after ignoring the ``duplicated" documents and considering only documents that cite just one BP each, we gathered 29,743 documents and 31,070 citations related to these 10 BPs, which corresponds to 77.5\% and 75.7\%, respectively, compared to all documents and citations from the initial dataset. 
Besides these 29,743 documents, we also consider other 30 thousand documents, without explicit citations, that we will use to identify potential citations (details in~\secref{sec:potentialCitation}).
Finally, our documents generally have less than four thousand words (we have a word count histogram in Appendix B).

%% file: 5-model.tex
\section{Identification of Potential Citations}
\label{sec:potentialCitation}

In the context of citations to binding precedents, relevant questions may arise.
For instance, \emph{``are there documents that potentially cite a BP?''} (\ie, the document should have cited the BP but it has not), or \emph{``are the document and the BP related enough?''} 
These questions can help understand how STF works and if the BPs are correctly applied.

In this section, we investigate the possibility of potential citations and how to find them. 
We describe our modeling of identification of potential citations as a classification task, and we discuss how to use classifiers to infer the citations. 
Our pipeline for identifying potential citations (\figref{fig:potential_pipeline}) is composed of two main steps: (a) the learning process, when the models learn what makes a citation, and (b) the potential citation inference, when we use the models to identify a potential citation and an interpretability technique to explain this decision. 
We give more details about these stages in the following.

\subsection{Learning Process}
\label{sec:learning_process}

Suppose there is a function $d_{X} \colon U \to \mathbb{R}$ that takes a document $D$ from our corpus $U$ and returns a score of citing the binding precedent $X$. When a document has a ``good'' score through $d_X$, we can assign it to a potential citation group. Motivated by this reasoning, the potential citation identification process starts searching for such a function $d_X$.

There are many ways to model the problem above. Two possible approaches are to consider the score $d_X$ as the distance between the document $D$ and the BP $X$ in some latent space or compute the probability of $D$ citing the BP $X$.
With these approaches, when the distance between $D$ and BP $X$ is small, or $D$ citing $X$'s probability is high, we say that we found a potential citation. 
The distance between $D$ and BP $X$ can be computed by embedding $D$ and BP $X$ in a high dimensional vector space (\eg, Doc2vec~\cite{doc2vec}) and then calculating the distance in this space. 
In the second case, we could compute the probability of a document $D$ citing BP $X$ by modeling the problem as a classification task. 
If we fit a classification model in documents that cite the BP $X$, the model can learn what makes a document cite the precedent. 
Therefore, we could employ the trained classifier to verify whether a new document cites the BP $X$, using the model's probability as a confidence level for this assignment. 

In this work, we rely on the classification approach to identify potential citations to BPs. 
The distance-based approach is similar to clustering with centers in the BPs, so the approach is unsupervised by nature.
On the other hand, the classification is supervised and leverages the labels of the other documents. It is also easier to identify a potential citation confidence threshold in terms of probabilities than distances (\eg, documents with model's probability greater than $t_c$ are assigned to a potential citation).
Moreover, it is more straightforward to apply interpretability techniques to classification models' decisions (\eg, using Lime \cite{lime} or Anchors \cite{ribeiro2018anchors}) instead of trying to interpret citation inference using vector representations.

\myparagraph{Data Pre-processing:}
Before training our models, we need to process our data and create a balanced sample dataset, further split into training, validation, and test datasets. 
We generate the sample dataset by choosing a random sample of documents that cite one of the ten most-cited BPs. 
As described in~\secref{sec:data_set}, we do not consider ``duplicated'' documents or documents citing more than one BP.
Avoiding duplicated documents is necessary to prevent the classifier from overfitting the data. 
Having documents citing only one BP ensures they only have one label, making the classification task more straightforward.

We chose only to consider the ten most-cited precedents. The reasons for this decision are twofold. 
First, the number of documents associated with less frequent BPs is not enough to train the models, demanding extra effort to remedy this situation, such as using a data augmentation scheme. 
Second, the reduced number of documents associated with the less frequent BPs can quickly be inspected manually, so there is no need to undertake efforts to develop sophisticated resources to analyze them.
In total, our sample dataset contains 6,730 documents balanced among the ten classes.

Given that most documents explicitly contain the text ``Binding Precedent \#X'' (in free translation), we remove these citations using regular expressions to avoid hints during the classification process.
The documents also underwent a standard NLP text pre-processing, which included converting the text to lowercase, \emph{tokenization}, removing punctuation and stop words, and \emph{lemmatization}.

\myparagraph{Text Embedding:}
As mentioned above, we use a classification model to assess the probability of a document citing a BP.
To assess different classifiers' performance, we rely on text embedding methods to generate a vector representation of the documents.
We test different embeddings to evaluate their performance: TF-IDF~\cite{Robertson2004}, Doc2vec~\cite{doc2vec}, Universal Sentence Encoder (USE)~\cite{use}, and Longformer~\cite{beltagy2020longformer}. 
In TF-IDF's particular case, we have also applied a dimensionality reduction procedure to map the documents to a 50-dimensional space using Truncated Singular Value Decomposition (SVD)~\cite{truncated_svd}. 
We call the final representation Truncated TF-IDF. 
The reason to consider Longformer, instead of a more popular language model, \eg BERT \cite{bert}, is that Longformer overcomes BERT's limitation of 512 tokens.
Note that USE and Longformer do not need a pre-processed text (as described in the previous paragraph) because both have their tokenization mechanism.

\figPipeline

\myparagraph{Classification:}
Support Vector Machines (SVM) are complex enough for real-world classification problems and simple enough to be analyzed mathematically~\cite{svm}. 
Our study considers SVM with linear and Radial Basis Function (RBF) kernels. Moreover, given that Longformer~\cite{beltagy2020longformer} has a linear classification neural network layer plugged into it, we also fine-tuned it to assess its classification performance in our context.

The described classifiers are used to search for potential citations in unlabeled data, assigning probabilities to each document to belong to each class. 
To get probabilities from an SVM, we need to make calibration: using labeled data, create a map from the classifier's output (SVM scores) to a probability estimate for each class, which sum up to 1.
To create this map, we use the calibration method from Platt et al.~\cite{platt1999probabilistic} for SVM with linear kernel and the Wu et al.~\cite{wu2004probability} method for RBF kernel.
Platt et al.'s method, precisely, does not support the multiclass case, so we calibrate each class in a ``one-vs-rest'' approach and normalize the results in the end to sum up to 1.
This map will receive the SVM scores from a data instance and output a probability estimate for each class.

\subsection{Potential Citation Inference}
\label{sec:pot-ite-inf}

This section describes how we use the classifiers to identify potential citations and understand the models' decisions (see~\figref{fig:potential_pipeline}~(b)).

\myparagraph{Citation Inference:}
The association between a text embedding technique and a classifier gives us what we call a \emph{model}. 
A model receives the raw text and returns the classifier's probabilities, which we interpret as the document's probability to cite each precedent. 
We consider a potential citation when the citation probability is greater or equal to a threshold $t_c \in [0,1]$ (chosen by the user).

\myparagraph{Interpretability:}
A concern when dealing with machine learning models is their interpretability. 
For instance, if Truncated TF-IDF combined with SVM points out that a document potentially cites BP 10 because of high returned probability, how can we understand the reasons behind this decision? Generally speaking, \emph{how can we trust this model?} 
Understanding why a model is taking a particular decision is of paramount importance~\cite{lime}, especially in sensitive scenarios like a legal document analysis.

Consider document $X$ and its embedding vector $\mathbf{x} \in \mathbb{R}^d$ (\eg, TF-IDF vector). 
This document's probability of belonging to class $C$ is given by $f_C(\mathbf{x})$, with $f_C \colon \mathbb{R}^d \to [0,1]$ a model's returned probability for class $C$. 
Our particular interest is to know the importance of each sentence from document $X$ to the given probability, \ie, if each sentence has positive, negative, or neutral importance, and the magnitude of this importance.

After preliminary tests have discarded the employment of the \emph{leave-one-out feature importance} (LOO)~\cite{loo} for interpretability due to its high sensitivity, we chose the \emph{Local Interpretable Model-agnostic Explanations} (Lime)~\cite{lime} method to tackle this issue.
The intuition is that, by removing a specific sentence and obtaining the variation of probability $\Delta f_C$, we can measure the importance of this sentence for the prediction $f_C(\mathbf{x})$. More formally, we randomly remove some sentences from document $X$, vectorize this new document to $\mathbf{z}$, and add it to set $Z$. Doing this many times, we have a collection $Z$ of vectors around $\mathbf{x}$ in high dimensional space.
We approximate $f_C(\mathbf{z})$ using a linear model $g(\mathbf{z})$, minimizing the approximation error $\mathcal{L}(f_C, g, \mathbf{z})$ in $\mathbf{z} \in Z$, but also constraining the complexity $\Omega(g)$ of $g$. 

This task can be interpreted as a weighted linear regression with regularization. We end with a model $g(\mathbf{z})$ that is linear over the sentences (\ie, the presence or absence of a sentence), where the linear model's coefficients can be interpreted as the importance score of each sentence to the final prediction $f_C(\mathbf{x})$.

By default, Lime works with words, and there is also the possibility of interpreting entire paragraphs. 
A word division brings some advantages, such as detailing. 
Still, it does not immediately assess each sentence's importance to the decision (\eg, sentences similar to the BP text). 
A paragraph division is also interesting for visualization purposes (see, for example, \secref{sec:document_reader}), but it merges various sentences into one, hampering the precise identification of relevant parts. 
Therefore, our choice of working with sentences comprises the ``best of both worlds''. In our experiments, Lime proved to be very robust and reliable when dealing with sentences.

\subsection{Modeling Results}
\label{sec:modeling_results}

The modeling described in \figref{fig:potential_pipeline} (\ie, fitting classifiers to data, using the classifiers to infer citations, and interpreting these inferences) needs to be validated. 
In this section, we evaluate the modeling process quantitatively and qualitatively.

\figConfusion

\myparagraph{Quantitative Analysis.}
We fit the classifiers with the text embedding to a training set corresponding to 80\% of our labeled, balanced sample dataset (randomly chosen, preserving balancing). Before applying SVM, each text embedding coordinate was standardized, improving SVM classification performance~\cite{standardizeSVM}. Then, we performed a grid search and cross-validation to optimize SVM parameters (linear and RBF). We also fine-tuned the Longformer model for four epochs, with a batch size of 12 and a linear decrease of the learning rate. In general, all models performed well in the validation set (10\% of the sample dataset, preserving balancing), especially TF-IDF-based models and Longformer, as depicted in \tabref{tab:accuracies}. In particular, Truncated TF-IDF with linear SVM has an excellent performance when predicting the correct BP on this set, as shown in the confusion matrix comparing the true with the predicted BPs from \figref{fig:confusion_matrix}.
Recall that the explicit mentions of the BPs were removed from the documents in the pre-processing step.

\begin{table}[t!]
  \color{black}
  \footnotesize
  \centering
  \caption{Evaluation of models' performance on validation portion of the sample dataset. Precision, recall, and F1 score are the average of each class' metrics weighted by number of class instances.}
  \label{tab:accuracies}
  \begin{tabular}{cc|cccc}
    \textbf{Embedding}                & \textbf{Classifier} & \textbf{Acc.} & \textbf{Precision} & \textbf{Recall} & \textbf{F1 score} \\ \hline
    \multirow{2}{*}{Trunc. TF-IDF} &  Linear          & 0.94              & 0.94               & 0.94            & 0.94              \\
                                      &  RBF             & 0.94              & 0.94               & 0.94            & 0.94              \\ \hline
    \multirow{2}{*}{TF-IDF}           &  Linear          & 0.94              & 0.94               & 0.94            & 0.94              \\
                                      &  RBF             & 0.93              & 0.93               & 0.93            & 0.93              \\ \hline
    \multicolumn{2}{c|}{Longformer}                         & 0.93              & 0.93               & 0.93            & 0.93              \\ \hline
    \multirow{2}{*}{Doc2vec}          &  Linear          & 0.84              & 0.85               & 0.84            & 0.84              \\
                                      &  RBF             & 0.88              & 0.88               & 0.88            & 0.88              \\ \hline
    \multirow{2}{*}{USE}              &  Linear          & 0.85              & 0.86               & 0.85            & 0.85              \\
                                      &  RBF             & 0.86              & 0.87               & 0.86            & 0.86              \\ \hline
  \end{tabular}
\end{table}

\myparagraph{Qualitative Analysis.}
The models' quantitative results support the idea that searching for potential citations can be solved as a classification task. 
Moreover, TF-IDF combined with SVM and Longformer presented pretty good results. 
However, Longformer tends to assign probability close to 1 to a particular class and the remaining probabilities close to zero. This fact makes it hard to interpret the probabilities returned by Longformer as a confidence level. 
In contrast, Truncated TF-IDF with SVM better distributes probabilities between 0 and 1. 
This fact added to other properties\,---\,ease of implementation, low dimensionality, and validation performance\,---\,rendered Truncated TF-IDF with linear SVM the better choice in our context.

\myparagraph{Chosen model on test data.}
The chosen model (Truncated TF-IDF with linear SVM), which had good performance on validation data (\autoref{tab:accuracies}), also performed well on test data (10\% of a sample dataset, balanced), achieving 0.96 in all metrics of \autoref{tab:accuracies}.

\myparagraph{Identification Analysis.}
To search for potential citations, we run our model with a document collection containing 30,000 documents that do not cite any BP (\secref{sec:data_set}). For sanity check, we also run our model in documents that explicitly cite the ten chosen BPs. 
Of course, when the model assigns a potential citation that is the same as the explicit citation, we continue calling it explicit citation. 
The goal is to rely on the classification model to find potential citations as one of the ten chosen BPs. 
The larger the user-defined threshold $t_c$ (\secref{sec:pot-ite-inf}, Citation Inference), the smaller the number of documents with potential citations, as presented in~\tabref{tab:potential_citations}.

\figVisualComp

\begin{table}[t!]
  \footnotesize
  \centering
  \caption{Number of documents with potential citations of each BP found by Truncated TF-IDF, for three different $t_c$ values.}
  
  \label{tab:potential_citations}
  \begin{tabularx}{\columnwidth}{XXXXXXXXXXX}
    \multirow{2}{*}{\textbf{$t_c$}} & \multicolumn{10}{c}{\textbf{BP}} \\
    & \textbf{3}  & \textbf{4}  & \textbf{10} & \textbf{11} & \textbf{14} & \textbf{17} & \textbf{20} & \textbf{26} & \textbf{33} & \textbf{37}\\ \hline

\textbf{0.99} & 4   & 73  & 1   & 1  & 2   & 7   & 4  & 5   & 50  & 1 \\ \hline
\textbf{0.95} & 45  & 237 & 65  & 8  & 78  & 184 & 59 & 118 & 344 & 42\\ \hline
\textbf{0.90} & 124 & 374 & 476 & 85 & 212 & 524 & 88 & 368 & 469 & 119\\ \hline
  \end{tabularx}
  
\end{table}
      
We use Lime (\secref{sec:pot-ite-inf}) to understand why the model assigned a potential citation to a document. 
The local model fitted by Lime is a linear classifier. We can interpret this model's weights as the importance of each sentence to the model's decision, \ie, the potential citation assignment. 
These weights were mapped into a color range in the document background to visually indicate which sentences are more important. 
The validation of this assignment's interpretability, so as the potential citation assignment, is established in the usage scenarios discussed in~\secref{sec:case_studies}.

%% file: 6-visual.tex
\section{LegalVis}
\label{sec:visual}

In this section we describe the visual components that compose LegalVis and also provide implementation details. As shown in~\figref{fig:visualComponents}, the system is made up of three main components, namely \textit{Global View}, \textit{Paragraph Similarities View}, and \textit{Document Reader}, that address the proposed design tasks as depicted in~\tabref{tab:views_tasks}.

\begin{table}[t!]
  \footnotesize
  \centering
  \caption{Visual components of LegalVis and the addressed tasks.}
  \label{tab:views_tasks}
  \begin{tabular}{lllllllllll}
    \multirow{2}{*}{} & \multicolumn{6}{c}{\textbf{Design Task}}                         \\
                                & \textbf{T1} & \textbf{T2} & \textbf{T3} & \textbf{T4} & \textbf{T5} & \textbf{T6} \\ \hline
    \textbf{Global View}                & \checkmark   & \checkmark   & \checkmark   &     &    &     \\ \hline
    \textbf{Parag. Similarities}                & \checkmark   &    &     & \checkmark    & \checkmark    &        \\ \hline
    \textbf{Document Reader}                & \checkmark   &    &     &     & \checkmark    & \checkmark 
  \end{tabular}
  % \vspace{-0.5cm}
\end{table}

\subsection{Global View}

\textit{Global View} (\figref{fig:visualComponents}~(a)) is the main and first opened panel. This view is responsible for showing an overview of the dataset under a temporal perspective (T2) and providing interactive tools that guide the user in searching for documents of interest (T3). Some documents in the dataset do not have valid date information, as will be discussed in~\secref{sec:limitations}. Since Global View depends on this attribute, these documents were potentially used to train the classification model (\secref{sec:potentialCitation}) but ignored in the visualization. 

In this view, the $x$-axis represents the time a document was published (monthly resolution), and the $y$-axis represents the BP. The red pins (\img{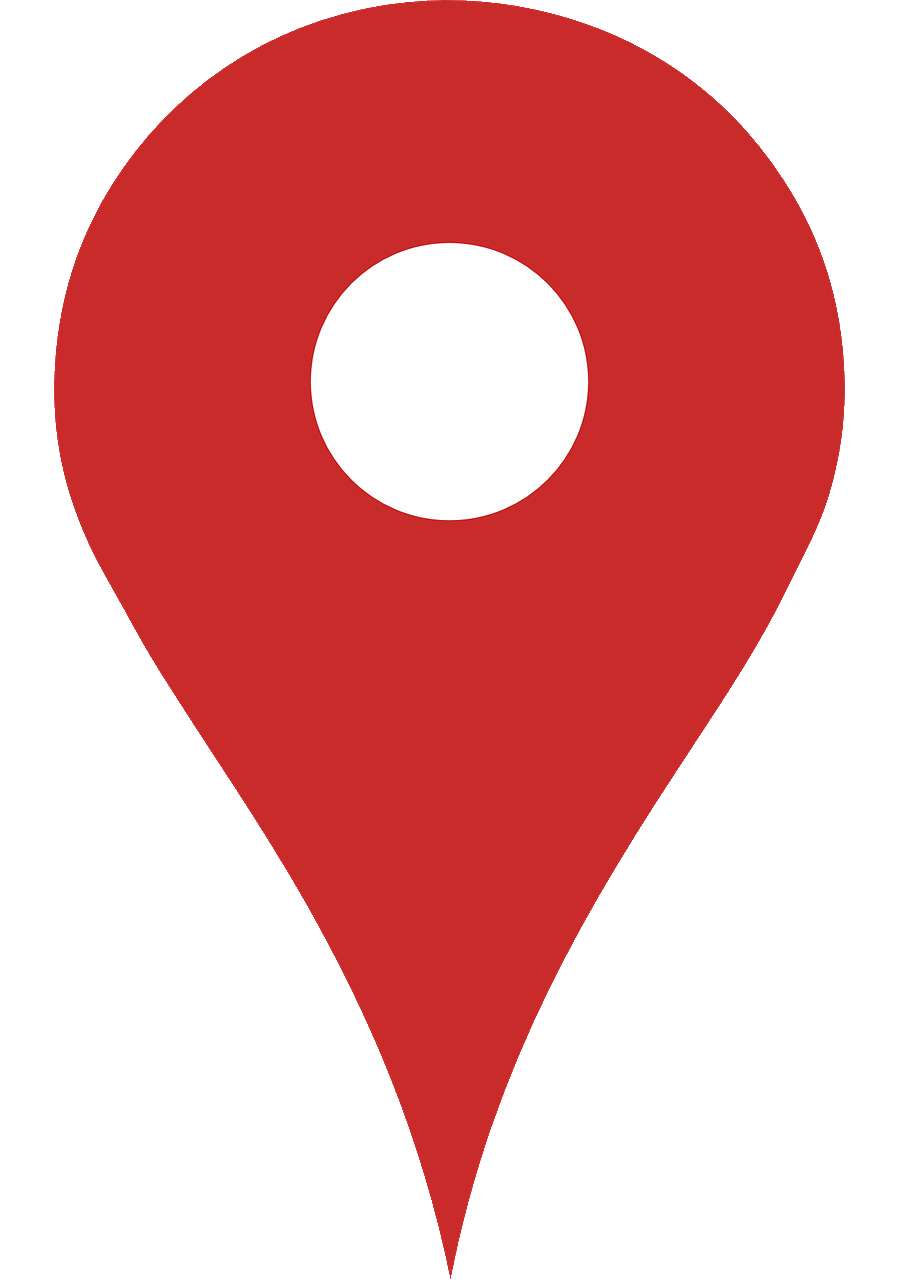}) refer to the publication date of the corresponding BPs. For each one, a vertical bar on a particular date indicates the existence of documents published on that date that cite such BP (T1). The height
of each bar reflects the number of documents published on that date, and its color is defined such that (i) blue bars (without borders) indicate 
that every document in that month cites the BP explicitly, (ii) orange bars (without borders) suggest that every document potentially (rather than explicitly) cites the BP, and (iii) blue bars with orange edges indicate the presence of both explicit and potential citations.

For better understanding and exploration of regions of interest, users can \textit{zoom~in}, \textit{zoom~out} and \textit{pan}. Other interactions include filtering documents by 
(i) the type of citation (explicit and/or potential); (ii) Justice rapporteur; (iii) document type (\eg, complaints or appeals); and 
(iv) potential citation confidence (threshold $t_c$). Finally, a tooltip showing the number of the BP, the date, and the number of documents (total, with explicit citations, and with potential citations) appears when a bar has hovered over. The publication date of the BP is also shown alongside the red pin.

We considered other design choices for \textit{Global View}, for example, fixed-size squares with the number of documents mapped by a color-scale (a strategy similar to the \textit{Temporal Activity Map}~\cite{TAM}) and varying-size circles instead of vertical bars. Since color coding was already considered the more suitable strategy to represent explicit/potential citations in this view (and in \textit{Paragraph Similarities View}, as we show in the next section), and given the high level of overlaps and cluttering obtained with varying-size circles, mapping the number of documents through bars with varying heights was preferred.

\subsection{Paragraph similarities View}

Once the user has found a set of documents (\ie, a bar) of interest in \textit{Global View}, he/she can select such a set by clicking on the bar. The user is then
redirected to the \textit{Paragraph Similarities View} (\figref{fig:visualComponents}~(b)), which presents in the $y$-axis all documents from the selected bar, that is,
documents that cite a particular BP on a given date. 

Each document is divided into paragraphs, indicated by the horizontal stack of bars. The bar's size means the size of the paragraph, and the color intensity reflects the similarity between the corresponding paragraph and the BP text; the darker the color, the greater the similarity, and the more common parts exist between the BP and the paragraph (T5). Similarly to Global View, the color of each stacked bar stands for explicit (blue) or potential (orange) citation (T1).
The similarity is given by the angular distance~\cite{use} between two Truncated-TF-IDF vectors, but other embeddings and similarities (\eg, the raw cosine similarity) could be adopted.

To guide users further exploring the set of documents, we group them into clusters (T4 -- details below). 
Inside each cluster, the documents are positioned in descending order of similarity between the document and the BP, 
which is defined as the maximum similarity between its paragraphs and the BP (T4). 
Showing documents in descending order of similarity is helpful because the user can promptly identify the top-$k$ most similar, and therefore most interesting, documents. Furthermore, to quickly assess the document distribution w.r.t. the similarity values, we also show an interactive bar chart above each color bar.

\myparagraph{Document Clustering.} Since a binding precedent may cover decisions related to different subjects, the Brazilian Supreme Court website provides, for each BP, some clusters of decisions created according to their subjects. There are only a few clusters for each BP, each containing a few labeled documents (for instance, there are 8 clusters for BP 4, each containing two documents on average). We employed NLP text pre-processing and applied topic
modeling strategies to cluster documents according to their relevant words to take advantage of this limited but useful ground truth. We have tested different and well-established topic extraction methods, including LDA~\cite{blei2003latent}, NMF~\cite{nmf}, SNMF~\cite{snmf}, and PSMF~\cite{psmf}. 
The NMF (Frobenius norm) method presented the best results, so we adopted it as the default method.

In the Paragraph Similarities View, users are free to choose the desired number of clusters/topics. 
A word cloud with the top-10 most relevant keywords is shown alongside each cluster to optimize the task of finding clusters of interest. Buttons \img{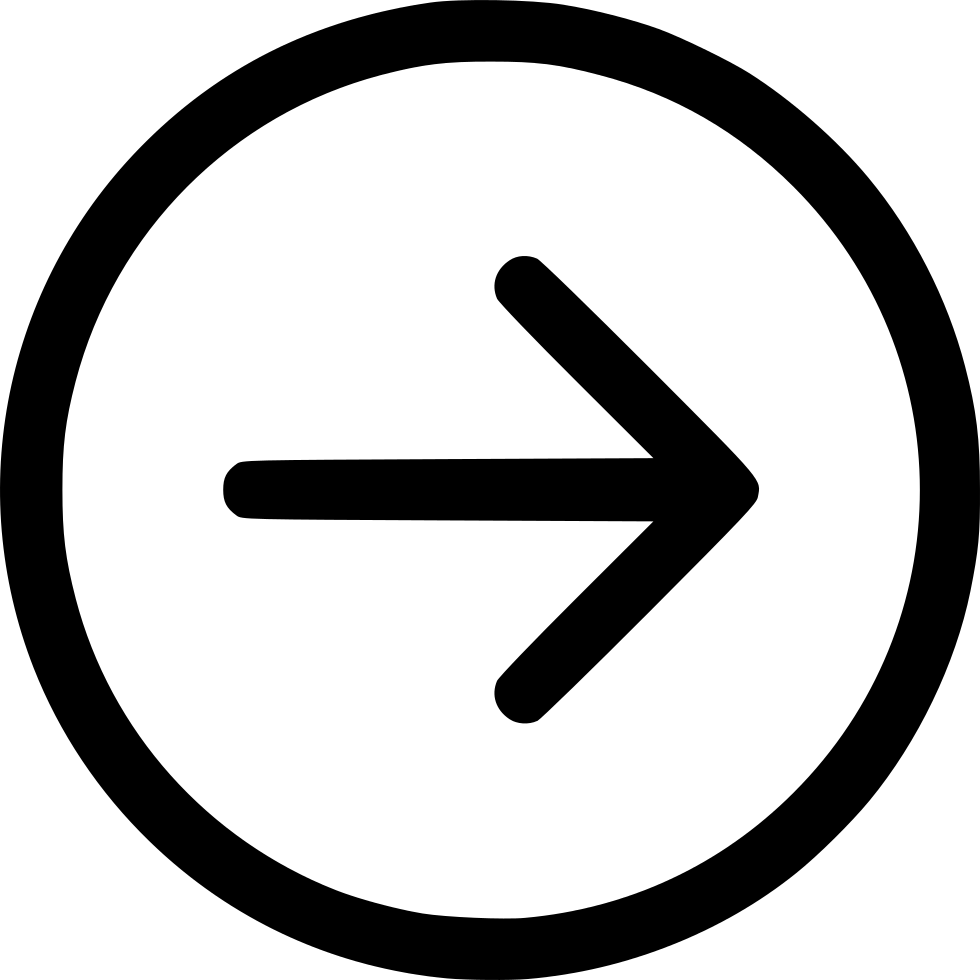} and \img{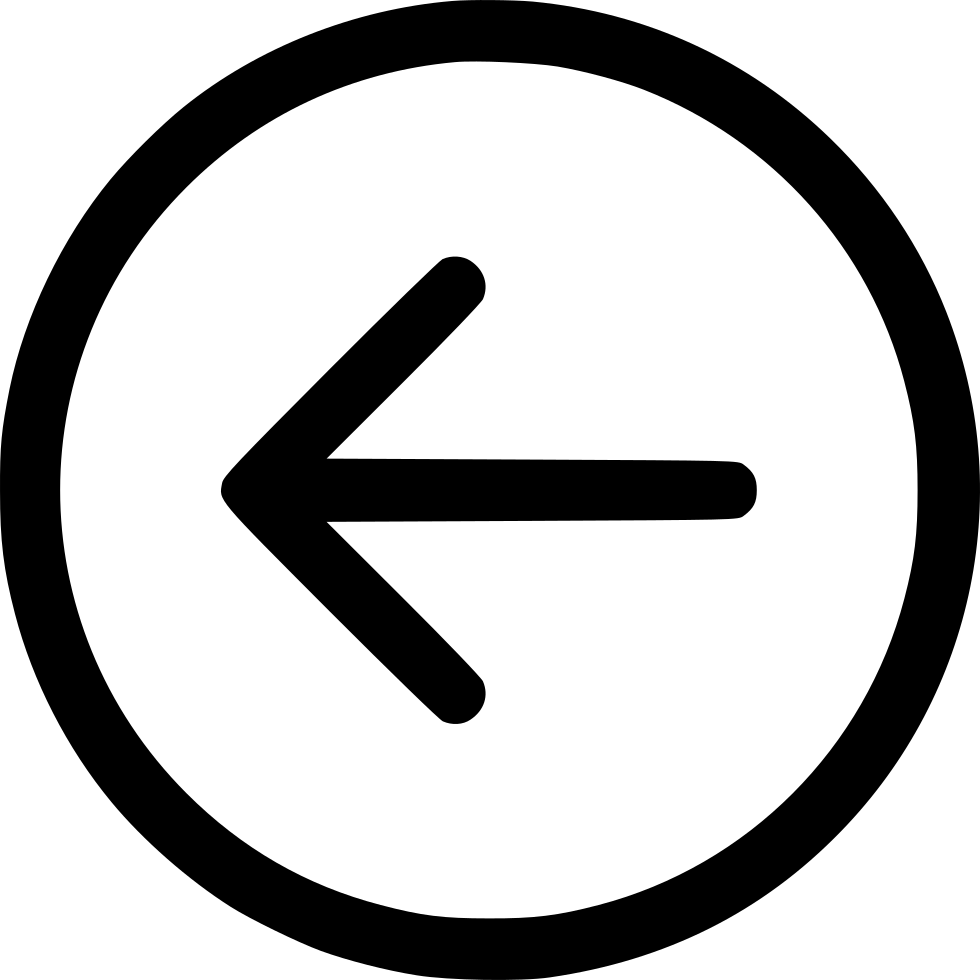} allow one to navigate between the detected clusters.

\subsection{Document Reader}
\label{sec:document_reader}

After finding a decision that seems interesting, the user selects one of its paragraphs by clicking on it. At this moment, he/she is redirected to \textit{Document Reader} (\figref{fig:visualComponents}~(c)), a view that enables the analysis of both the document's content and the BP text. In this view, the colored bars from the Paragraph Similarities View are also visible alongside the paragraphs (\figref{fig:visualComponents}~(c) and \figref{fig:show_text_similarity}). 

When a paragraph is hovered over, its content and the BP text are shown as illustrated in~\figref{fig:show_text_similarity}, with the common parts highlighted or not depending on the user's needs (T5 --- flag ``\textit{Show text similarity}''). In the example shown in the figure, one may notice that the text of the paragraph associated with the darkest blue bar is contained in the BP text (see yellow highlights). To analyze any other paragraph of the current document, one may hover over its text without revisiting Paragraph Similarities View. 
 
Document Reader also incorporates Lime when the opened document refers to a potential citation (T1, T5). In this case, instead of just showing the document's raw text (as in \figref{fig:show_text_similarity}), the parts of the document's content are highlighted according to their influence (negative, neutral, or positive) in the potential citation identification process (see~\figref{fig:visualComponents}~(c)).

\figShowTextSimilarity

Since one may analyze many documents published on different dates and citing different BPs, Document Reader also keeps track of recently opened documents (T6), 
allowing the user to quickly revisit them through the document browsing history (\figref{fig:show_text_similarity}). 

In practice, it is not uncommon to find complex cases in which a single justice writes his or her decision in a document with more than 100 pages.
Document Reader minimizes experts' reading time by tracking recently opened documents of interest and pointing out exactly which paragraphs of the document are likely to mention the binding precedent, even if not explicitly. These characteristics increase the time efficiency of those who search for specific citations in long documents.

\subsection{Implementation Details}
\label{sec:implementation_details}

Our system adopts a client-server architecture. From the client perspective (front-end), all three visual components were implemented using the D3 library~\cite{d3js}. The server-side (back-end) was implemented in Python and employs some well-established libraries and tools, such as Flask~\cite{flask}, Scikit-learn~\cite{scikit-learn}, NLTK~\cite{nltk}, ElasticSearch~\cite{elastic}, and others.
For storing and retrieving documents' and binding precedents' information, LegalVis relies on a MySQL database.

%% file: 7-cases.tex
\figCaseOne

\section{Usage Scenarios}
\label{sec:case_studies}

In this section, we discuss two usage scenarios that demonstrate the capabilities of LegalVis in exploring legal documents related to binding precedents. The first usage scenario is related to tasks T1, T2, T5 and highlights the system's effectiveness in identifying potential citations under three different situations. The second usage scenario addresses T2, T3, T4, and T6 by showing an exploratory analysis involving related decisions reported by two particular justice rapporteurs. Unless explicitly
stated otherwise, all analyses adopt $t_c = 0.95$. To better understand, we also translated those parts of documents and binding precedents crucial for the analyses to English. The original text contents (in Portuguese) are available in Appendix A.

\subsection{Exploring Potential Citations}

As observed in the explicit citations, potential citations are expected to exist in documents published after the publication of the binding precedent they refer to. 
There are, though, documents with potential citations to a particular BP whose publishing date is before the creation date of that BP. 
Recall that a binding precedent represents a consolidated understanding among the Supreme Court's justices. Thus a series of decisions may have existed 
related to this underlying judicial issue that led to the precedent's creation. In this usage scenario, we explore three particular 
potential citations (\figref{fig:caseStudy1}), two of them identified in documents prior to the binding precedent publication (\figref{fig:caseStudy1},~left~and~middle).

The first potential citation involves a document published in Feb. 2013 containing an instrument appeal (\textit{``agravo de instrumento''}, in free translation). According to the system, this document potentially cites BP~37, even though this BP was published in Oct. 2014 (\figref{fig:caseStudy1}~(a)). When analyzing how similar the paragraphs of this document are to BP~37 in the Paragraph Similarities View, we see that one of them is very close to it (the darkest orange bar in~\figref{fig:caseStudy1}~(b)). By looking at the content of this presumable relevant paragraph in the Document Reader (\figref{fig:caseStudy1}~(c)), one may notice that Lime considered a long part of it as of great importance in the citation identification process (see the darkest green background on the text). Recall that Document Reader also allows users to compare the text of the corresponding binding precedent with any selected paragraph, with the possibility of highlighting their common parts. By using this feature (\figref{fig:caseStudy1}~(d)), we can easily perceive that this paragraph contains the entire BP text, with a single modification (``on the'' \textit{vs} ``on''). At this point, a relevant question emerges: \textit{How can a document have the text of a BP before its creation?} In fact, this document does not quote BP 37; instead, it quotes Precedent 339 (\textit{Súmula 339}), a precedent that was eventually converted into BP 37 \cite{sv37} (see explicit mention to Precedent 339 just before the highlighted parts in~\figref{fig:caseStudy1}~(d)).

A potential citation with the explicit BP text is an exception, especially for documents published before the creation of the BP. The expected behavior is that documents and BPs involved in potential citations have the same implicit reasoning. Our second analysis (\figref{fig:caseStudy1}~(middle)) illustrates this case by considering an instrument appeal, published in Nov. 2004 (\figref{fig:caseStudy1}~(e)), that potentially cites BP 4 (created in April 2008). After selecting one of its paragraphs in the
Paragraph Similarities View (\figref{fig:caseStudy1}~(f)), we can compare its text with the BP text in the Document Reader.
When analyzing the BP text and also two parts that were considered relevant by Lime in two adjacent paragraphs (the selected one and a neighbor), we see that both the document and the BP refer to the prohibition of using the minimum wage as an indexing factor (\figref{fig:caseStudy1}~(g)), which supports the claim of a valid potential citation; they however describe this subject in different ways.

The two cases described so far demonstrate the usefulness of LegalVis for identifying and analyzing documents published before creating the binding precedent they are somehow related to. Identifying and exploring these documents help users to understand why the Supreme Court creates binding precedents, one of the key needs pointed out by judicial experts that LegalVis successfully addresses through tasks T1 (identification of potential citations), T2 (overview of documents under a temporal perspective), and T5 (highlight in relevant parts of the documents).

In practice, a justice may refer to a binding precedent X as ``binding entry \#X of the precedent" (\textit{``verbete vinculante no X da súmula''}, in free translation). This latter terminology was not considered when labeling the documents, so (i) documents adopting it were marked as not having explicit citations (\ie, unlabeled documents), and (ii) it can be used as ground-truth to validate the identification process. To study this case, we rely on a complaint (\textit{``reclamação''}) that potentially cites BP 10 when adopting $t_c = 0.91$ (\figref{fig:caseStudy1}~(right)). This document was selected among all documents that (explicitly or potentially) cite BP 10 in Sep. 2013 (\figref{fig:caseStudy1}~(h-i)). When comparing the text of the BP with one of the paragraphs marked as of great importance by Lime (\figref{fig:caseStudy1}~(j)), we can notice that both refer to some violation of the so-called plenary reserve clause. Still, while the BP defines the rule, the document brings a real-world and related situation. Although this common subject suggests a coherent and valid potential citation, we can validate it by observing an explicit mention to ``Binding Entry \#10 of the Precedent of the Supreme Court" in another document's paragraph (\figref{fig:caseStudy1}~(j)). This example supports once more the usefulness of LegalVis in identifying and exploring potential citations and demonstrates the potential of the system to optimize and validate document labeling.

\subsection{Related Decisions}

Users may explore LegalVis to search for patterns involving decisions and/or BPs. As shown in~\figref{fig:caseStudy2}, this usage scenario presents an exploratory analysis focusing on related decisions reported by two Supreme Court's justices, Teori Zavascki and Edson Fachin. 

\figCaseTwo

BP 14 covers a criminal law theme. It grants lawyers the right to access every documented evidence already collected during an investigation that could help them prove their client's innocence. It was applied, for example, at the \textit{Lava Jato} case, a criminal investigation started in 2014 in Brazil on alleged irregularities involving several individuals and companies, among them Petrobras, the largest state-owned company in Brazil \cite{lavajato}.

Among the 64 documents of type ``petition'' (\textit{``petição''}, in Portuguese) that cite or are likely to cite BP 14, 33 are associated with Teori Zavascki and 21 to Edson Fachin. 
Likewise, among the 67 documents of type ``investigation'' (\textit{``inquérito''}) related to this BP, Zavascki is the rapporteur of 26, and Fachin has 18, followed by other justices. Except for a single document, Fachin only started dealing with documents of these types after Zavascki stopped, as illustrated by \figref{fig:caseStudy2}~(a) for documents ``investigation''. 
A brief background is required to understand this behavior: Teori Zavascki was the Supreme Court's justice rapporteur for the Lava Jato operation cases. Several of Zavascki's decisions, therefore, were related to Lava Jato. He passed away in Jan. 2017 due to an airplane crash, and Fachin was appointed as a new rapporteur for the Lava Jato's cases from that moment on \cite{lavajatonovorelatorstf}.

In this usage scenario, rather than validating potential citation identification, we want to find out if there is a relation between decisions from Zavascki and Fachin connected to Lava Jato. We begin our exploratory analysis by selecting in Global View all documents associated with Edson Fachin for BP~14 in Sep. 2017. The five retrieved documents are shown in Paragraph Similarities View as a single cluster. After asking for two clusters using topic modeling, the documents are clustered as shown in~\figref{fig:caseStudy2}~(b), \ie, a cluster ``Topic~1'' containing two complaints (documents of type ``Rcl'') that explicitly cite BP 14 and a cluster ``Topic~2'', containing three investigations (documents of type ``Inq'') that potentially cite BP 14. We also see terms related to investigation (\eg, ``investigar'' and ``apurar'') and ``petrobras'' in the word-cloud relative to Topic~2 (\figref{fig:caseStudy2}~(c)), which is interesting for our search. It is worth noting that the quality of this clustering supports the suitability of topic modeling for separating the documents. As expected, documents of different types tend to be characterized by different words, and consequently, separated into different clusters.

By exploring Topic 2's last document (``Inq 4413'') with Document Reader, we can identify three parts in the text that are relevant for our analysis (\figref{fig:caseStudy2}~(d)). By observing these parts, we can finally establish the relationship we were looking for: Fachin explicitly agrees with Zavascki's decision at ``Investigation 4,231'' (\figref{fig:caseStudy2}~(d-2)). This decision, which is related to Lava Jato and Petrobras (\figref{fig:caseStudy2}~(d-3)), is unfortunately under legal secrecy and is not available for analysis. In his decision, Fachin also mentions two other investigations (ids 4,325 and 4,326 -- see \figref{fig:caseStudy2}~(d-1)), where one may find other agreements or controversies. By identifying this kind of relation between decisions or justices, experts can understand and explore the adopted arguments in new and related processes or anticipate possible outcomes of ongoing ones. The system also facilitates comparisons of decisions. If we open multiple documents in the system (\eg, those related to these investigations), we can promptly switch between them by using the document browsing history. 

Recall that LegalVis has identified a potential citation between ``Inq 4413'' and BP 14 with confidence $t_c = 0.95$. Even though we could not establish reasons that justify this identification, the presented usage scenario showed the system's effectiveness in assisting users in finding patterns and behaviors related to particular justices and documents' types. Since LegalVis properly tackles tasks T2 and T3, finding the temporal relationship between cases reported by Zavasci and Fachin was easy in the explored scenario. Reaching the analyzed document of interest was easy, mainly because of the topic modeling clustering (T4). Finally, comparisons among different decisions are possible and uncomplicated through the document browsing history~(T6).

%% file: 8-experts.tex
\section{Evaluation from Experts}
\label{sec:experts}

This section describes an evaluation with experts performed to collect detailed feedback about the presented usage scenarios, the usefulness and usability of the tool, and ideas for further improvement.

\subsection{Participants}

We recruited six domain experts not involved in the tool's development to evaluate our proposal. They work as attorneys (2), researchers (2), legal assistants (1), and trainees (1) and have from 1.5 to 22 years of experience in analyzing legal documents and/or binding precedents. The participation was voluntary and without payment.

\subsection{Evaluation Process}
The expert evaluation involved three steps. First, the participants had to watch a video presenting the pipeline to identify potential citations, the visual components of LegalVis, and the two usage scenarios described in \secref{sec:case_studies}. After that, we invited them to explore the tool through two well-defined tasks:

\myparagraph{T1:} We asked the participants to select the documents associated with Justice Ricardo Lewandowski and: (i) identify the BP least cited by him; (ii) identify the time interval in which he did not cite any BP (if any); (iii) based on the provided word cloud, give their opinion on what they think the 16 decisions associated to him and explicitly citing BP 20 on May 2014 refer to; (iv) find a decision, reported by him, that not only explicitly cites a BP (any BP) but also contains the BP's text. While we consider (i) and (ii) relatively straightforward as they can be answered through Global View with few interactions (filtering options), our goal with (iii) was to guide the participants to Paragraph Similarities View and evaluate how helpful and trustful the word cloud containing relevant keywords is. We also aimed to assess the system for exploring decisions with explicit citations (iv).

\myparagraph{T2:} We asked the participants to find any decision with a potential citation ($t_c = 0.95$) and inform which part of the decision's text they consider relevant w.r.t. the corresponding BP. Besides evaluating the system for explorations involving potential citations, we aimed to assess the quality of the results found by LegalVis on this matter.

Finally, the experts answered a set of quantitative (QT) and qualitative (QL) questions. Regarding the quantitative ones, participants had to answer, through a 5-point Likert scale, whether they agree with the following statements: 
``\textit{Usage scenario 1 is relevant.}'' (QT1);
``\textit{Usage scenario 2 is relevant.}'' (QT2);
``\textit{It is easy to find decisions with explicit or potential citations throughout time.}'' (QT3);
``\textit{It is easy to filter interesting decisions for analysis.}'' (QT4); 
``\textit{It is easy to identify parts of decisions related to the BP of interest.}'' (QT5);
``\textit{LegalVis is useful.}'' (QT6); 
``\textit{It is easy to learn how to use LegalVis.}'' (QT7); 
``\textit{LegalVis is efficient and would optimize my time.}'' (QT8); 
``\textit{LegalVis is easy to use.}'' (QT9); 
``\textit{LegalVis has an intuitive interface.}'' (QT10).

Besides asking for comments on QT3-QT6, and also a general comment on QT7-QT10, the qualitative questions include: ``\textit{What is your impression about the proposed pipeline for the identification of potential citations?}'' (QL1); 
``\textit{What do you highlight as relevant or interesting in usage scenarios 1 (QL2) and 2 (QL3)? }'';
``\textit{Without LegalVis, what are the challenges in performing analysis similar to the ones you have made?}'' (QL4); 
``\textit{What other tools that allow this type of analysis do you know?}'' (QL5); 
``\textit{What are the advantages (QL6) and disadvantages (QL7) of LegalVis compared to other tools you are used to?}''; 
``\textit{In your opinion, which are the most helpful visual components?}'' (QL8); 
``\textit{Besides the existing visual components, which other components do you think could be incorporated in LegalVis?}'' (QL9); 
``\textit{Would you like to leave a final comment?}'' (QL10). We only asked QL6 and Q7 to those who know other tools that allow similar analyses (QL5).

\subsection{Results}

Five of the six participants responded to T1(i) correctly, and all six provided similar/correct answers for T1(ii-iv). We invalidated one answer for T2 because the participant used a document with explicit citation (instead of a potential one) to answer it; four of the other participants provided answers in line with Lime.
We present in the following the experts' opinions about (i) the pipeline of identification of potential citations, (ii) relevance of the usage scenarios described in~\secref{sec:case_studies}, (iii) system's usefulness, (iv) usability, and (v) scope for further improvements.

\myparagraph{Potential citation identification (QL1, QL10):} 
The experts considered the pipeline interesting, robust, and promising. One of the participants commented: \emph{``Although I don't know how the machine learning works, the proposal is exciting and, without doubt, of great value not only to lawyers but also to all careers that depend on the analysis of judicial decisions."} For another expert, \emph{``The pipeline seems excellent. The use and improvement of machine learning algorithms lead to great results"}. According to a third participant, \emph{``It enhances the search for binding precedents and similar decisions, which helps attorneys to support their petitions"}.

\myparagraph{Usage scenarios (QT1, QT2, QL2, QL3):} 
The first usage scenario was considered relevant by five of the six participants (\figref{fig:likert}~(QT1)). According to their comments, they were particularly delighted because it \textit{``clearly demonstrates the applicability of the tool''} and its results show that LegalVis \textit{``enables and facilitates the identification of previous court positions that led to the creation of BPs''} in a \textit{``practical and intelligent way''}. 

\figLikert

Although the overall rating for the second usage scenario is also positive (\figref{fig:likert}~(QT2)), one participant considered it not relevant. He/she stated: \textit{``This analysis involves going through several steps that seem less necessary. If the objective is to compare decisions, a simpler way would be to compare the decisions they cite and their content directly''}. It is worth noting that this usage scenario describes the comparison between decisions and demonstrates how users can find relevant decisions and justices to be compared. For another participant, this usage scenario is relevant because it shows that the system \textit{``helps to find similar decisions, which bring unity to the judiciary''}. A third expert commented: \textit{``I foresee great applicability in the field of legal arguments. Great.''}

\myparagraph{System's usefulness (QT6, QT8, QL4-QL6, QL10):}
All participants consider LegalVis a useful tool, and all but one think it is efficient and time-saving (\figref{fig:likert}~(QT6, QT8)). One of them commented: \textit{``While learning the features, I already anticipated using the tool better illustrate my classes, opinions, and scientific articles''}. According to another participant: \textit{``It will greatly help all law professionals, bringing more quality to the services provided and reducing research time''}. A third expert commented: \textit{``I hope I can use the tool for professional purposes as soon as possible.''}

According to the experts, LegalVis is useful because it addresses some critical issues related to large-scale data analysis (3 participants mentioned this issue), filtering options (2), and searches based only on keywords (3). Only two participants responded `yes' when asked whether they know similar tools. The tools they mentioned are the STF's website and a platform named JusBrasil~\cite{jusbrasil}. According to one participant, \textit{``the search offered by these websites is based only on keywords and some filters. Thus, LegalVis proves to be much more complete and useful for the better development of activities related to legal arguments. In my opinion, the LegalVis proposal is broader and more useful.''}
Another participant commented: \textit{``LegalVis seems to have a proposal that simplifies large-scale data analysis. Most jurists I know rely on search tools based on not much more than keywords. In that sense, LegalVis would be an important tool''}. A third expert highlighted one problem with keyword-based search engines: \textit{``Without LegalVis, we would have to search for keywords on the STF's website. The problem is that it is not always possible to think of good keywords to use. Sometimes even their search engine is poor.''} 

Besides the advantages mentioned above, LegalVis was also considered useful because \textit{``potential citations seem to be an excellent source of argument to use in the Courts.''} This participant's comment continues: \textit{``I consider the tool very useful, not only in terms of litigation but also in the development of academic activities (scientific articles, research, classes, etc.).''}

\myparagraph{System's usability (QT3-QT5, QT7, QT9, QT10, QL8):}
Five of the six participants consider it easy to find decisions with citations throughout time (\figref{fig:likert}~(QT3)), especially those with explicit citations. They also agree that one can quickly identify parts of decisions related to the BP of interest (\figref{fig:likert}~(QT5)). Filtering interesting decisions for analysis, on the other hand, was considered a more difficult task (\figref{fig:likert}~(QT4)). About that, one participant commented: \textit{``The learning curve of the system is a little slower at first due to the complexity and a large number of utilities that can be extracted. However, I emphasize that it is possible to filter very interesting documents for analysis''}. Despite this comment about the learning curve, most experts agree that it is easy to learn how to use LegalVis (\figref{fig:likert}~(QT7)). 

According to most participants, LegalVis is easy to use and has an intuitive interface (\figref{fig:likert}~(QT9, QT10)). Regarding the most helpful visual components, the experts reported different opinions: \textit{``the similarity color scale''}, \textit{``highlight in relevant excerpts''}, \textit{``timeline visualization''}, \textit{``similarity bar alongside the decision's paragraph''}, \textit{``all of them are useful''}.

\myparagraph{Scope for improvements (QT4, QT10, QL7, QL9, QL10):}
Although LegalVis shows strong capabilities to enhance the analysis of legal documents and binding precedents, it can still be improved as suggested by the domain experts. 
Regarding filtering interesting decisions and new features, received suggestions include \textit{``filtering by justice and document type are interesting, but I think more important is a keyword filter, similar to the one offered by the STF's website, to make it easier to search for a specific subject.''} and \textit{``it would be nice if LegalVis could incorporate subject filtering''}. It is worth mentioning that, since we have both the decisions' texts and the capability to detect relevant keywords through topic modeling, LegalVis could incorporate these features with little effort. 

Some experts felt that LegalVis could become more user-friendly and offer more filtering possibilities. One of them commented: \textit{``The system is really cool! I think, however, that it is still possible to improve the visual aspect, for example, by removing from the visualization those BPs that are not subject of the person's research''}. Another expert also suggested filtering decisions based on time intervals defined by the user. LegalVis can naturally incorporate these filtering options.

About the quality of the potential citation identification, one participant reported that \textit{``there seem to be a few cases where the potential citation is not very related to the BP''}. We will discuss ideas for improving our model in the next section.

%% file: 9-discussion.tex
\section{Discussion and Limitations}
\label{sec:limitations}

The presented usage scenarios and experts' evaluations show that LegalVis is a helpful tool to assist judicial experts in analyzing legal documents. 
LegalVis can also be adapted to assist experts from the legal field in deciding whether or not a BP should be associated with a process under analysis, reducing the processing times. 
A similar framework could also be applied to other domains like medicine and public purchases. Medical prescriptions and government purchases can be seen as legal documents while the regulatory rules as BPs, for example. 
Another potential application is in scientific research, where papers to be cited can be seen as precedents for new ones.

\myparagraph{TF-IDF performance.} An important question from our results is why TF-IDF presents better performance than the new generation of text representations tools (Doc2vec, USE, and Longformer).
One hypothesis is that the presence of specific words dictates the behavior of the classifier (SVM in our case). We performed several experiments considering both a decision tree classifier over bag-of-words and a logistic regression using TF-IDF with L1 regularization in a one-vs-rest multiclass classification approach to analyze this aspect further. Both experiments showed that the presence of specific words led to high accuracy in the classification  (over 85\% on validation set).
However, each classification method highlights a different set of relevant words, making it hard to establish a consensus on which words are essential for classifying each BP.
This fact reinforces the critical role of Lime, as it allows for breaking the document into sentences rather than words, facilitating the interpretation. Moreover, Lime explanations are local (per sample), enabling individual understanding of the model's decision.

The issue related to the inferior performance of Doc2vec, USE, and Longformer may also be associated with the lack of pre-trained models with Brazilian Portuguese, in particular with Brazilian Portuguese legal language, which has several particularities in the writing style and terms used throughout the documents. This key factor might be negatively impacting the models' performance.
Training models to suit the characteristics of legal data is a task we are very interested in, and the accomplishment of this task is in our near future plans.
It is essential to mention that in certain cases, TF-IDF can, indeed, outperform Doc2vec, as already reported by some previous work~\cite{7822730, Wang_2017, tfidf_melhor_alguns_datasets}.

\myparagraph{Problems with data.} The dataset we use came from a partnership with another project (recall~\secref{sec:data_set}). Unfortunately, it presents some limitations that required some adaptations in our system: (i) presence of some ``duplicated'' documents, \ie, documents with different titles but same raw texts that had to be ignored in the classification to avoid overfitting the data; (ii) presence of some documents without valid date attribute (marked as Jan. 1970), which were ignored in the visualization;
(iii) presence of some documents for which it was impossible to automatically extract justice information (marked as ``unknown justice'').

\myparagraph{Potential citation identification model.} Although efficient, our model is simple in the sense that it only considers the documents' raw texts. As future work, we intend to incorporate metadata (\eg, justice rapporteur and date) directly into the process. These metadata already exist as parts of the texts, but we hypothesize that giving more weight to them would improve the identification. We also plan to experiment with other models (\eg, \textit{Big Bird}~\cite{zaheer2020big}) and analyze their performance. A promising experiment would be to test with Transformer-based models pretrained using Portuguese texts (\eg, BERTimbau~\cite{souza2020bertimbau}), differently of Longformer, which we hypothesize is a downside in our case. Not least, the system's current version does not allow users to influence the model's decisions. We now intend to aggregate user relevance feedback to improve the performance through additional training. Such feedback would be especially relevant when dealing with more (and imbalanced) BPs or citations to multiple BPs.

\myparagraph{Number of documents and binding precedents.} There are 58 binding precedents up to this moment, and we chose to explore only the ten most cited of them. Before increasing this number, one should be aware that there are binding precedents with just a few citations, consequently impairing the classification task. To address these issues, we intend to explore data augmentation methods (\eg, EDA~\cite{wei2019eda}) and other unbalanced classification methods \cite{krawczyk_learning_2016}. We also plan to exhibit each class performance and each document's potential citation probability. This way, users are better informed and can influence the system's decisions through the relevance feedback mentioned.

\myparagraph{Topic Modeling.} As shown in the second usage scenario, our clustering strategy based on topic modeling is helpful to guide users to documents of interest. However, the lack of ground truth with a representative sample of decisions impaired a formal evaluation of its quality. Although validated by domain experts for particular cases, we intend to augment our ground truth and perform a more robust analysis.

%% file: 10-conclusion.tex
\section{Conclusion}

In this paper, we presented LegalVis, a web-based visual analytic system designed to assist lawyers and other judicial experts in analyzing legal documents that cite or could potentially cite binding precedents. LegalVis first identifies potential citations by implementing a simple yet powerful machine learning model based on classification. Then, an interpretability mechanism also incorporated into the system provides reliable explanations for the model's decisions. Finally, all this information becomes accessible through the three interactive and linked views that compose the system. Qualitative and quantitative analyses validated the performance of the proposed model and two usage scenarios demonstrated the usefulness and effectiveness of LegalVis system. Both scenarios were validated by six domain experts not involved in the system's development, who also reported positive feedback.

%% file: paper.bbl
\begin{thebibliography}{10}

\bibitem{abdul2017constructive}
A.~Abdul-Rahman, G.~Roe, M.~Olsen, C.~Gladstone, R.~Whaling, N.~Cronk,
  R.~Morrissey, and M.~Chen.
\newblock Constructive visual analytics for text similarity detection.
\newblock {\em Computer Graphics Forum}, 36(1):237--248, 2017.

\bibitem{alharbi2018sos}
M.~Alharbi and R.~S. Laramee.
\newblock {SoS TextVis}: A survey of surveys on text visualization.
\newblock In {\em Conference on Computer Graphics \& Visual Computing}, p.
  143–152. Eurographics Association, Goslar, DEU, 2018.

\bibitem{standardizeSVM}
S.~Ali and K.~A. Smith-Miles.
\newblock Improved support vector machine generalization using normalized input
  space.
\newblock In A.~Sattar and B.-h. Kang, eds., {\em AI 2006: Advances in
  Artificial Intelligence}, pp. 362--371. Springer Berlin Heidelberg, Berlin,
  Heidelberg, 2006.

\bibitem{alsakran2011streamit}
J.~Alsakran, Y.~Chen, Y.~Zhao, J.~Yang, and D.~Luo.
\newblock {STREAMIT}: Dynamic visualization and interactive exploration of text
  streams.
\newblock In {\em Pacific Visualization Symposium (PacificVis)}, p. 131–138.
  IEEE, USA, 2011.

\bibitem{tfidf_melhor_alguns_datasets}
M.~Avinash and E.~Sivasankar.
\newblock A study of feature extraction techniques for sentiment analysis.
\newblock In A.~Abraham, P.~Dutta, J.~K. Mandal, A.~Bhattacharya, and S.~Dutta,
  eds., {\em Emerging Technologies in Data Mining and Information Security},
  pp. 475--486. Springer Singapore, Singapore, 2019.

\bibitem{barros2018case}
R.~Barros, A.~Peres, F.~Lorenzi, L.~Krug~Wives, and E.~Hubert~da
  Silva~Jaccottet.
\newblock Case law analysis with machine learning in {B}razilian court.
\newblock In M.~Mouhoub, S.~Sadaoui, O.~Ait~Mohamed, and M.~Ali, eds., {\em
  Recent Trends and Future Technology in Applied Intelligence}, pp. 857--868.
  Springer International Publishing, Cham, 2018.

\bibitem{beltagy2020longformer}
I.~Beltagy, M.~E. Peters, and A.~Cohan.
\newblock Longformer: The long-document transformer.
\newblock {\em CoRR}, abs/2004.05150, 2020.

\bibitem{nltk}
S.~Bird, E.~Klein, and E.~Loper.
\newblock {\em Natural language processing with {P}ython: analyzing text with
  the natural language toolkit}.
\newblock "O'Reilly Media, Inc.", 2009.

\bibitem{blei2003latent}
D.~M. Blei, A.~Y. Ng, and M.~I. Jordan.
\newblock Latent dirichlet allocation.
\newblock {\em Journal of Machine Learning Research}, 3:993--1022, 2003.

\bibitem{bluetick}
{Bluetick}.
\newblock {Bluetick | Vinden zonder zoeken} {[Bluetick | Find without searching
  (free translation)]}.
\newblock \url{https://www.bluetick.nl/}. Accessed: 2021-09-06.

\bibitem{cnjpendingcases}
{{Brazilian National Council of Justice}}.
\newblock {Número total de casos pendentes na justiça brasileira [Total
  number of pending cases in the Brazilian legal system (free translation)]}.
\newblock
  \url{https://www.cnj.jus.br/wp-content/uploads/2013/01/fc587d0cef86f585d6872a4f0ff43107.zip}.
  Accessed: 2021-03-30.

\bibitem{stfnumberrulings}
{Brazilian Supreme Court}.
\newblock {Número de decisões do STF entre 2011 e 2020 [Number of rulings on
  STF between 2011 and 2020 (free translation)]}.
\newblock
  \url{http://www.stf.jus.br/portal/cms/verTexto.asp?servico=estatistica&pagina=decisoesgeral}.
  Accessed: 2021-03-30.

\bibitem{stfsumulas}
{{Brazilian Supreme Court}}.
\newblock {S\'umulas Vinculantes [Binding Precedents]}.
\newblock
  \url{http://www.stf.jus.br/portal/cms/verTexto.asp?servico=jurisprudenciaSumulaVinculante}.
  Accessed: 2021-03-30.

\bibitem{sv37}
{{Brazilian Supreme Court}}.
\newblock {Aplicação das Súmulas no STF - S\'umula Vinculante 37
  [Application of STF Precedents - Binding Precedent \#37 (free translation)]},
  2014.
\newblock
  \url{http://www.stf.jus.br/portal/jurisprudencia/menuSumario.asp?sumula=1961}.
  Accessed: 2021-03-30.

\bibitem{lavajatonovorelatorstf}
{{Brazilian Supreme Court}}.
\newblock {Ministro Edson Fachin é sorteado novo relator da Lava-Jato [Justice
  Edson Fachin is appointed as new rapporteur of cases involving the Car Wash
  operation (free translation)]}, 2017.
\newblock
  \url{http://www.stf.jus.br/portal/cms/verNoticiaDetalhe.asp?idConteudo=335003}.
  Accessed: 2021-03-30.

\bibitem{dizerDireito}
{Buscador Dizer o Direito}.
\newblock {Buscador Dizer o Direito: Encontre jurisprudência comentada do STF
  e do STJ e muito mais} {[Search engine Saying the Law: Find commented
  jurisprudence of the STF and STJ and much more (free translation)]}.
\newblock \url{https://www.buscadordizerodireito.com.br/}. Accessed:
  2021-09-06.

\bibitem{cao2016overview}
N.~Cao and W.~Cui.
\newblock Overview of text visualization techniques.
\newblock In {\em Introduction to Text Visualization. Atlantis Briefs in
  Artificial Intelligence}, vol.~1. Atlantis Press, 2016.

\bibitem{carvalho2018transforming}
N.~R. Carvalho and L.~S. Barbosa.
\newblock Transforming legal documents for visualization and analysis.
\newblock In {\em 11th International Conference on Theory and Practice of
  Electronic Governance}, p. 23–26. ACM, Galway, Ireland, 2018.

\bibitem{casetext}
{Casetext}.
\newblock {Compose: Craft exceptional briefs, without the busy work}.
\newblock \url{https://compose.law/}. Accessed: 2021-09-06.

\bibitem{use}
D.~Cer, Y.~Yang, S.~Kong, N.~Hua, N.~Limtiaco, R.~S. John, N.~Constant,
  M.~Guajardo{-}Cespedes, S.~Yuan, C.~Tar, Y.~Sung, B.~Strope, and R.~Kurzweil.
\newblock {Universal Sentence Encoder}.
\newblock {\em CoRR}, abs/1803.11175, 2018.

\bibitem{chada2015visualizing}
D.~M. Chada, F.~A. Silva, and P.~Borges.
\newblock Visualizing {Brazilian Justice: The Supreme Court} 2.0 project.
\newblock In {\em 15th International Conference on Artificial Intelligence and
  Law}, p. 176–180. ACM, San Diego, California, 2015.

\bibitem{cheema2016annotatevis}
M.~F. Cheema, S.~J{\"a}nicke, and G.~Scheuermann.
\newblock {AnnotateVis}: Combining traditional close reading with visual text
  analysis.
\newblock In {\em Workshop on Visualization for the Digital Humanities, IEEE
  VIS}. Baltimore, Maryland, USA, 2016.

\bibitem{nmf}
A.~Cichocki and A.-H. Phan.
\newblock Fast local algorithms for large scale nonnegative matrix and tensor
  factorizations.
\newblock {\em {IEICE Transactions on Fundamentals of Electronics,
  Communications and Computer Sciences}}, 92(3):708--721, 2009.

\bibitem{correia2019exploratory}
F.~A. Correia, J.~L. Nunes, G.~F. C.~F. de~Almeida, A.~A.~A. Almeida, and
  H.~Lopes.
\newblock An exploratory analysis of precedent relevance in the {Brazilian
  Supreme Court} rulings.
\newblock In {\em Symposium on Document Engineering}. ACM, Berlin, Germany,
  2019.

\bibitem{d3js}
{D3.js}.
\newblock {D3.js - Data-Driven Documents}.
\newblock \url{https://d3js.org}. Accessed: 2021-09-06.

\bibitem{bert}
J.~Devlin, M.~Chang, K.~Lee, and K.~Toutanova.
\newblock {BERT:} pre-training of deep bidirectional transformers for language
  understanding.
\newblock {\em CoRR}, abs/1810.04805, 2018.

\bibitem{diakopoulos2015compare}
N.~Diakopoulos, D.~Elgesem, A.~Salway, A.~Zhang, and K.~Hofland.
\newblock Compare clouds: Visualizing text corpora to compare media frames.
\newblock In {\em IUI Workshop on Visual Text Analytics}, pp. 193--202.
  Atlanta, GA, USA, 2015.

\bibitem{psmf}
D.~Dueck, Q.~D. Morris, and B.~J. Frey.
\newblock Multi-way clustering of microarray data using probabilistic sparse
  matrix factorization.
\newblock {\em Bioinformatics}, 21(suppl\_1):i144--i151, 2005.

\bibitem{elastic}
{Elastic}.
\newblock {Elasticsearch: The official distributed search \& analytics engine}.
\newblock \url{https://www.elastic.co/elasticsearch/}. Accessed: 2021-09-06.

\bibitem{falcao2013relatorio}
J.~Falc{\~a}o, P.~Cerdeira, and D.~Arguelhes.
\newblock I relat{\'o}rio do supremo em n{\'u}meros-o m{\'u}ltiplo supremo [1st
  supreme in numbers report - the supreme multiple (free translation)].
\newblock {\em Revista de Direito Administrativo}, 262:399--452, 2013.

\bibitem{finchplatform}
{Finch Platform}.
\newblock {Finch - Simplificando o mundo jurídico} {[Finch - Simplifying the
  legal world (free translation)]}.
\newblock \url{https://finchsolucoes.com.br/}. Accessed: 2021-09-06.

\bibitem{gomez2015understanding}
E.~Gomez-Nieto, W.~Casaca, I.~Hartmann, and L.~G. Nonato.
\newblock Understanding large legal datasets through visual analytics.
\newblock In {\em 6th Workshop on Visual Analytics, Information Visualization
  and Scientific Visualization (WVIS) in SIBGRAPI}, vol.~15. Salvador, Brazil,
  2015.

\bibitem{flask}
M.~Grinberg.
\newblock {\em Flask web development: developing web applications with
  {P}ython}.
\newblock "O'Reilly Media, Inc.", 2018.

\bibitem{truncated_svd}
N.~Halko, P.~G. Martinsson, and J.~A. Tropp.
\newblock Finding structure with randomness: Probabilistic algorithms for
  constructing approximate matrix decompositions.
\newblock {\em SIAM Review}, 53(2):217--288, 2011.

\bibitem{svm}
M.~Hearst, S.~Dumais, E.~Osuna, J.~Platt, and B.~Sch\"olkopf.
\newblock Support vector machines.
\newblock {\em IEEE Intelligent Systems and their Applications}, 13(4):18--28,
  1998. doi: {{%
10\hspace{.1pt}\discretionary{.}{%
}{.}\hspace{.4pt}1109\discretionary{/}{%
}{/}5254\hspace{.1pt}\discretionary{.}{%
}{.}\hspace{.4pt}708428}}


\bibitem{janicke2017visual}
S.~J{\"a}nicke, G.~Franzini, M.~F. Cheema, and G.~Scheuermann.
\newblock Visual text analysis in digital humanities.
\newblock {\em Computer Graphics Forum}, 36(6):226--250, 2017.

\bibitem{janicke2014visualizations}
S.~J{\"a}nicke, A.~Ge{\ss}ner, M.~B{\"u}chler, and G.~Scheuermann.
\newblock Visualizations for text re-use.
\newblock In {\em International Conference on Information Visualization Theory
  and Applications (IVAPP)}, pp. 59--70. IEEE, Lisbon, Portugal, 2014.

\bibitem{jusbrasil}
{Jusbrasil}.
\newblock {Jusbrasil. Conectando pessoas à Justiça} {[Jusbrasil. Connecting
  people to Justice (free translation)]}.
\newblock \url{https://www.jusbrasil.com.br/home}. Accessed: 2021-09-06.

\bibitem{kiesel2020visual}
D.~Kiesel, P.~Riehmann, H.~Wachsmuth, B.~Stein, and B.~Froehlich.
\newblock Visual analysis of argumentation in essays.
\newblock {\em IEEE Transactions on Visualization and Computer Graphics},
  27(2):1139--1148, 2021.

\bibitem{snmf}
H.~Kim and H.~Park.
\newblock Sparse non-negative matrix factorizations via alternating
  non-negativity-constrained least squares for microarray data analysis.
\newblock {\em Bioinformatics}, 23(12):1495--1502, 2007.

\bibitem{krawczyk_learning_2016}
B.~Krawczyk.
\newblock Learning from imbalanced data: open challenges and future directions.
\newblock {\em Progress in Artificial Intelligence}, 5(4):221--232, 2016.

\bibitem{kucher2015text}
K.~Kucher and A.~Kerren.
\newblock Text visualization techniques: Taxonomy, visual survey, and community
  insights.
\newblock In {\em Pacific Visualization Symposium (PacificVis)}, pp. 117--121.
  IEEE, Los Alamitos, CA, USA, 2015.

\bibitem{wyner2017answering}
D.~Kuppevelt and G.~Dijck.
\newblock Answering legal research questions about {Dutch} case law with
  network analysis and visualization.
\newblock In A.~Wyner and G.~Casini, eds., {\em Legal Knowledge and Information
  Systems: JURIX 2017: The Thirtieth Annual Conference}, vol. 302 of {\em
  Frontiers in Artificial Intelligence and Applications}, pp. 95--100. IOS
  Press, 2017.

\bibitem{doc2vec}
Q.~Le and T.~Mikolov.
\newblock Distributed representations of sentences and documents.
\newblock In E.~P. Xing and T.~Jebara, eds., {\em 31st International Conference
  on Machine Learning}, vol.~32 of {\em Proceedings of Machine Learning
  Research}, pp. 1188--1196. PMLR, Bejing, China, 2014.

\bibitem{lettieri2016computational}
N.~Lettieri, A.~Altamura, A.~Faggiano, and D.~Malandrino.
\newblock A computational approach for the experimental study of {EU} case law:
  analysis and implementation.
\newblock {\em Social Network Analysis and Mining}, 6(1):56, 2016.

\bibitem{lettieri2017legal}
N.~Lettieri, A.~Altamura, and D.~Malandrino.
\newblock The legal macroscope: Experimenting with visual legal analytics.
\newblock {\em Information Visualization}, 16(4):332--345, 2017.

\bibitem{lexis}
{Lexis}.
\newblock {Lexis - Online legal research}.
\newblock \url{https://www.lexisnexis.com/en-us/products/lexis.page}. Accessed:
  2021-09-06.

\bibitem{lexml}
{LexML}.
\newblock {LexML Brasil: Rede de informação legislativa e jurídica} {[LexML
  Brazil: Network of legislative and legal information (free translation)]}.
\newblock \url{https://www.lexml.gov.br}. Accessed: 2021-09-06.

\bibitem{loo}
J.~Li, W.~Monroe, and D.~Jurafsky.
\newblock Understanding neural networks through representation erasure.
\newblock {\em CoRR}, abs/1612.08220, 2016.

\bibitem{TAM}
C.~D.~G. Linhares, J.~R. Ponciano, J.~G.~S. Paiva, B.~A.~N. Traven{\c{c}}olo,
  and L.~E.~C. Rocha.
\newblock Visualisation of structure and processes on temporal networks.
\newblock In P.~Holme and J.~Saram{\"a}ki, eds., {\em Temporal Network Theory},
  pp. 83--105. Springer International Publishing, Cham, 2019.

\bibitem{legal_document_review}
C.~J. Mahoney, J.~Zhang, N.~Huber-Fliflet, P.~Gronvall, and H.~Zhao.
\newblock A framework for explainable text classification in legal document
  review.
\newblock In {\em International Conference on Big Data}, pp. 1858--1867. IEEE,
  Los Alamitos, CA, USA, 2019.

\bibitem{mandal2017measuring}
A.~Mandal, R.~Chaki, S.~Saha, K.~Ghosh, A.~Pal, and S.~Ghosh.
\newblock Measuring similarity among legal court case documents.
\newblock In {\em India Compute Conference}, p. 1–9. ACM, Bhopal, India,
  2017.

\bibitem{oabjuris}
{OABJuris}.
\newblock {OABJuris | Jurisprudência de uma forma mais ágil e eficaz}
  {[OABJuris | Jurisprudence in a more agile and effective way (free
  translation)]}.
\newblock \url{https://jurisprudencia.oab.org.br/}. Accessed: 2021-09-06.

\bibitem{scikit-learn}
F.~Pedregosa, G.~Varoquaux, A.~Gramfort, V.~Michel, B.~Thirion, O.~Grisel,
  M.~Blondel, P.~Prettenhofer, R.~Weiss, V.~Dubourg, J.~Vanderplas, A.~Passos,
  D.~Cournapeau, M.~Brucher, M.~Perrot, and E.~Duchesnay.
\newblock Scikit-learn: Machine learning in {P}ython.
\newblock {\em Journal of Machine Learning Research}, 12:2825--2830, 2011.

\bibitem{platt1999probabilistic}
J.~C. Platt.
\newblock Probabilistic outputs for support vector machines and comparisons to
  regularized likelihood methods.
\newblock In A.~J. Smola, P.~Bartlett, B.~Schölkopf, and D.~Schuurmans, eds.,
  {\em Advances in Large Margin Classifiers}, Neural Information Processing
  series, pp. 61--74. MIT Press, 1999.

\bibitem{potthast2013overview}
M.~Potthast, M.~Hagen, T.~Gollub, M.~Tippmann, J.~Kiesel, P.~Rosso,
  E.~Stamatatos, and B.~Stein.
\newblock Overview of the 5th international competition on plagiarism
  detection.
\newblock In {\em CLEF Conference on Multilingual and Multimodal Information
  Access Evaluation}, pp. 301--331. CELCT, Valencia, Spain, 2013.

\bibitem{lime}
M.~T. Ribeiro, S.~Singh, and C.~Guestrin.
\newblock "{Why} should {I} trust you?": Explaining the predictions of any
  classifier.
\newblock In {\em International Conference on Knowledge Discovery and Data
  Mining}, p. 1135–1144. ACM, San Francisco, USA, 2016.

\bibitem{ribeiro2018anchors}
M.~T. Ribeiro, S.~Singh, and C.~Guestrin.
\newblock Anchors: High-precision model-agnostic explanations.
\newblock In {\em 32nd AAAI Conference on Artificial Intelligence}, vol.~32,
  pp. 1527--1535. AAAI Press, New Orleans, USA, 2018.

\bibitem{riehmann2015visual}
P.~Riehmann, M.~Potthast, B.~Stein, and B.~Froehlich.
\newblock Visual assessment of alleged plagiarism cases.
\newblock {\em Computer Graphics Forum}, 34(3):61--70, 2015.

\bibitem{Robertson2004}
S.~Robertson.
\newblock Understanding inverse document frequency: on theoretical arguments
  for {IDF}.
\newblock {\em Journal of Documentation}, 60(5):503--520, 2004.

\bibitem{visualmantra}
B.~Shneiderman.
\newblock The eyes have it: A task by data type taxonomy for information
  visualizations.
\newblock In {\em Symposium on Visual Languages}, pp. 336--343. IEEE, USA,
  1996.

\bibitem{soto2015similarity}
A.~J. Soto, A.~Mohammad, A.~Albert, A.~Islam, E.~Milios, M.~Doyle, R.~Minghim,
  and M.~C. Ferreira~de Oliveira.
\newblock Similarity-based support for text reuse in technical writing.
\newblock In {\em Symposium on Document Engineering}, p. 97–106. ACM, New
  York, NY, USA, 2015.

\bibitem{souza2020bertimbau}
F.~Souza, R.~Nogueira, and R.~Lotufo.
\newblock {BERTimbau: Pretrained BERT models for Brazilian Portuguese"}.
\newblock In R.~Cerri and R.~C. Prati, eds., {\em Intelligent Systems. BRACIS
  2020. Lecture Notes in Computer Science}, vol. 12319, pp. 403--417. Springer
  International Publishing, Cham, 2020.

\bibitem{legalone}
{Thomson Reuters}.
\newblock {Legal One, a solução jurídica que se adequa à sua realidade}
  {[Legal One, the legal solution that suits your reality (free translation)]}.
\newblock \url{https://www.thomsonreuters.com.br/pt/juridico/legal-one.html}.
  Accessed: 2021-09-06.

\bibitem{westlaw}
{Thomson Reuters}.
\newblock {Westlaw - Legal research tools \& platforms}.
\newblock \url{https://legal.thomsonreuters.com/en/products/westlaw}. Accessed:
  2021-09-06.

\bibitem{van2008visualizing}
L.~Van~der Maaten and G.~Hinton.
\newblock Visualizing data using {t-SNE}.
\newblock {\em Journal of Machine Learning Research}, 9(11):2579–2605, 2008.

\bibitem{transformer}
A.~Vaswani, N.~Shazeer, N.~Parmar, J.~Uszkoreit, L.~Jones, A.~N. Gomez,
  L.~Kaiser, and I.~Polosukhin.
\newblock Attention is all you need.
\newblock {\em CoRR}, abs/1706.03762, 2017.

\bibitem{Wang_2017}
Y.~Wang, Z.~Zhou, S.~Jin, D.~Liu, and M.~Lu.
\newblock Comparisons and selections of features and classifiers for short text
  classification.
\newblock {\em {IOP} Conference Series: Materials Science and Engineering},
  261:012018, 2017.

\bibitem{lavajato}
J.~Watts.
\newblock {Operation Car Wash: Is this the biggest corruption scandal in
  history?}, 2017.
\newblock
  \url{https://www.theguardian.com/world/2017/jun/01/brazil-operation-car-wash-is-this-the-biggest-corruption-scandal-in-history}.
  Accessed: 2021-03-30.

\bibitem{wei2019eda}
J.~Wei and K.~Zou.
\newblock {EDA}: Easy data augmentation techniques for boosting performance on
  text classification tasks.
\newblock In {\em Conference on Empirical Methods in Natural Language
  Processing and the 9th International Joint Conference on Natural Language
  Processing (EMNLP-IJCNLP)}, pp. 6382--6388. Association for Computational
  Linguistics, Hong Kong, China, 2019.

\bibitem{wu2004probability}
T.-F. Wu, C.-J. Lin, and R.~C. Weng.
\newblock Probability estimates for multi-class classification by pairwise
  coupling.
\newblock {\em Journal of Machine Learning Research}, 5:975--1005, 2004.

\bibitem{surveyTextAlignment}
T.~Yousef and S.~J{\"a}nicke.
\newblock A survey of text alignment visualization.
\newblock {\em IEEE Transactions on Visualization and Computer Graphics},
  27(2):1149--1159, 2021. doi: {{%
10\hspace{.1pt}\discretionary{.}{%
}{.}\hspace{.4pt}1109\discretionary{/}{%
}{/}TVCG\hspace{.1pt}\discretionary{.}{%
}{.}\hspace{.4pt}2020\hspace{.1pt}\discretionary{.}{%
}{.}\hspace{.4pt}3028975}}


\bibitem{zaheer2020big}
M.~Zaheer, G.~Guruganesh, A.~Dubey, J.~Ainslie, C.~Alberti,
  S.~Onta{\~{n}}{\'{o}}n, P.~Pham, A.~Ravula, Q.~Wang, L.~Yang, and A.~Ahmed.
\newblock {Big Bird}: Transformers for longer sequences.
\newblock {\em CoRR}, abs/2007.14062, 2020.

\bibitem{zhang2007semantics}
P.~Zhang and L.~Koppaka.
\newblock Semantics-based legal citation network.
\newblock In {\em 11th International Conference on Artificial Intelligence and
  Law}, p. 123–130. ACM, California, USA, 2007.

\bibitem{7822730}
W.~Zhu, W.~Zhang, G.-Z. Li, C.~He, and L.~Zhang.
\newblock A study of damp-heat syndrome classification using {Word2vec and
  TF-IDF}.
\newblock In {\em International Conference on Bioinformatics and Biomedicine
  (BIBM)}, pp. 1415--1420. IEEE, Shenzhen, China, 2016.

\end{thebibliography}
